\documentclass[aps,onecolumn,nofootinbib,superscriptaddress,showkeys]{revtex4} 

\usepackage{epsfig}
\usepackage{graphicx}

\usepackage[utf8]{inputenc}

\usepackage{latexsym,amsmath,amssymb,amsfonts}
\usepackage{bbm}
\usepackage{graphics}

\usepackage{color}
\definecolor{linkcolor}{rgb}{0.4,0.1,0.1}
\definecolor{bibcolor}{rgb}{0.4,0.1,0.1}

  \long\def\comment#1{ }

\def\empile#1\over#2{\mathrel{\mathop{\kern 0pt#1}\limits_{#2}}}
\newcommand{\slvarepsilon}{\raise.15ex\hbox{$/$}\kern-.53em\hbox{$\varepsilon$}}
\newcommand{\slL}{\raise.15ex\hbox{$/$}\kern-.53em\hbox{$L$}}
\newcommand{\slP}{\raise.15ex\hbox{$/$}\kern-.53em\hbox{$P$}}
\newcommand{\slp}{\raise.1ex\hbox{$/$}\kern-.63em\hbox{$p$}}
\newcommand{\slq}{\raise.1ex\hbox{$/$}\kern-.53em\hbox{$q$}}
\newcommand{\slv}{\raise.1ex\hbox{$/$}\kern-.63em\hbox{$v$}}
\newcommand{\slR}{\raise.15ex\hbox{$/$}\kern-.53em\hbox{$R$}}
\newcommand{\slQ}{\raise.15ex\hbox{$/$}\kern-.53em\hbox{$Q$}}
\newcommand{\slK}{\raise.15ex\hbox{$/$}\kern-.53em\hbox{$K$}}
\newcommand{\slk}{\raise.15ex\hbox{$/$}\kern-.53em\hbox{$k$}}
\newcommand{\slSigma}{\raise.15ex\hbox{$/$}\kern-.53em\hbox{$\Sigma$}}
\newcommand{\slcalP}{\raise.15ex\hbox{$/$}\kern-.63em\hbox{$\cal P$}}
\newcommand{\slA}{\raise.15ex\hbox{$/$}\kern-.73em\hbox{$A$}}
\newcommand{\slbfA}{\raise.15ex\hbox{$/$}\kern-.73em\hbox{${\imb A}$}}
\newcommand{\slpartial}{\raise.15ex\hbox{$/$}\kern-.53em\hbox{$\partial$}}
\newcommand{\sla}{\raise.15ex\hbox{$/$}\kern-.53em\hbox{$a$}}
\newcommand{\slb}{\raise.15ex\hbox{$/$}\kern-.53em\hbox{$b$}}
\newcommand{\slc}{\raise.15ex\hbox{$/$}\kern-.53em\hbox{$c$}}
\newcommand{\slC}{\raise.15ex\hbox{$/$}\kern-.63em\hbox{$C$}}

\def\p{{\boldsymbol p}}

\def\k{{\boldsymbol k}}

\def\x{{\boldsymbol x}}
\def\v{{\boldsymbol v}}
\def\u{{\boldsymbol u}}

\def\X{{\boldsymbol X}}
\def\P{{\boldsymbol P}}
\def\r{{\boldsymbol r}}
\def\s{{\boldsymbol s}}

\def\bs{\boldsymbol}

\def\colora{}
\def\colorb{}
\def\colorc{}
\def\colord{}

  \newcommand{\beq}{\begin{eqnarray}}
  \newcommand{\eeq}{\end{eqnarray}}
  
 \def\simge{\mathrel{%
   \rlap{\raise 0.511ex \hbox{$>$}}{\lower 0.511ex \hbox{$\sim$}}}}
\def\simle{\mathrel{
   \rlap{\raise 0.511ex \hbox{$<$}}{\lower 0.511ex \hbox{$\sim$}}}}

\newcommand{\ud}{\, \mathrm{d}}

\newcommand{\rt}{{\r_\perp}}

\newcommand{\xt}{{\x_\perp}}

\begin{document}





\title{Initial State Quantum Fluctuations in the Little Bang}

\author{Fran\c cois Gelis}
\affiliation{Institut de physique th\'eorique, CEA, CNRS, Universit\'e Paris-Saclay, F-91191 Gif-sur-Yvette, France}
\author{Bj\"orn Schenke}
\affiliation{Physics Department, Brookhaven National Laboratory, Upton, NY 11973, USA}

\begin{abstract}
  We review recent developments in the ab-initio theoretical
  description of the initial state in heavy-ion collisions.  We
  emphasize the importance of fluctuations, both for the
  phenomenological description of experimental data from the
  Relativistic Heavy Ion Collider (RHIC) and the Large Hadron Collider
  (LHC), and the theoretical understanding of the non-equilibrium
  early time dynamics and thermalization of the medium.
\end{abstract}


\maketitle

\section{Introduction}
Heavy Ion Collisions, devoted to probe the hot and dense phases of
nuclear matter, may in principle be described by Quantum
Chromodynamics (QCD) -- the microscopic theory of quarks and
gluons. However, due to the extremely complicated dynamics at play in
the collisions, such an ab initio description appears
hopeless. Instead, one resorts to various mesoscopic descriptions,
that integrate out many of the microscopic details.

A crucial challenge in such a coarse graining procedure is to identify
the relevant aspects of the underlying fundamental description that
must be kept in order to correctly describe the system. In the
underlying quantum field theory, renormalizability implies that
quantum fluctuations are important down to spatial scales of the order
of the typical inverse momentum, while smaller fluctuations are simply
encapsulated in the scale dependence of a few parameters such as the
coupling constant. But in the more macroscopic descriptions used in heavy ion collisions there 
is no such clear procedure. 

The purpose of this review is to discuss the various sources of fluctuations, and
their role in the observable outcome of the collisions.

\begin{figure}[htbp]
\begin{center}
\includegraphics[width=0.8 \textwidth]{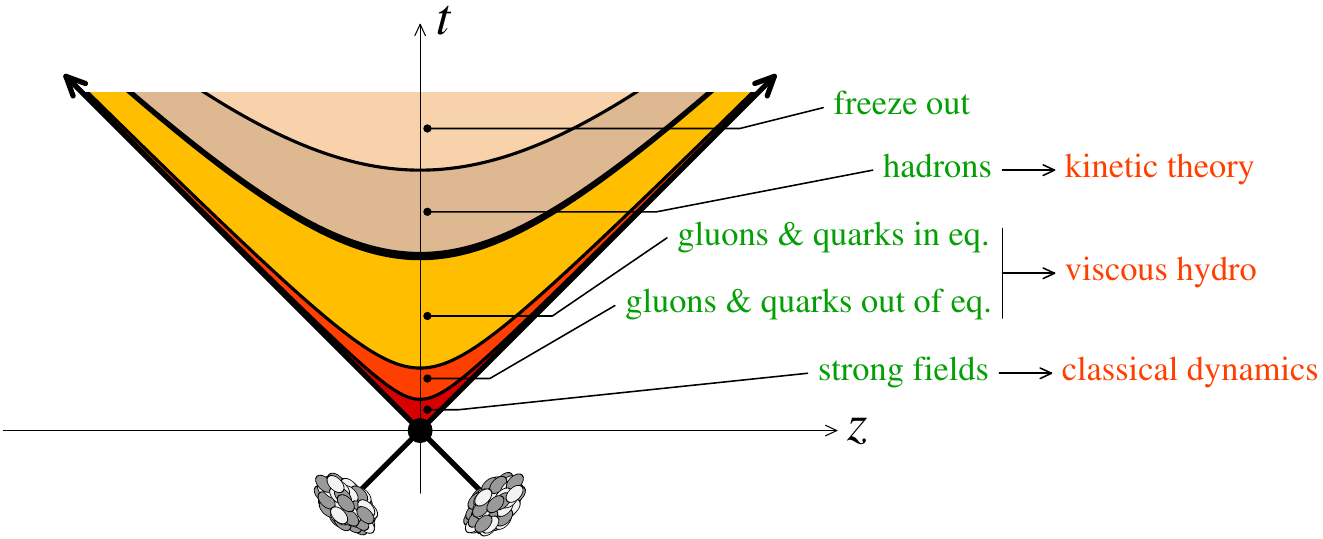}
\end{center}
\caption{\label{fig:stages}Successive stages of a heavy ion collision.}
\end{figure}

\subsection{Hydrodynamics in heavy ion collisions}
Over many years, relativistic fluid dynamics has proven to be the most
successful effective theory to describe the bulk dynamics of heavy ion
collisions.  Early on, ideal hydrodynamics was able to describe many
qualitative features of the experimental data, including a large
elliptic flow $v_2$ and mass splitting between the $v_2$ coefficients
for different particle species. This gave the first indication that
the matter created in heavy ion collisions at the Relativistic Heavy
Ion Collider (RHIC) is close to a perfect fluid
\cite{Hirano:2002ds,Kolb:2003dz,Huovinen:2003fa}. This was confirmed later at the
Large Hadron Collider (LHC).

With the development of viscous relativistic hydrodynamic simulations and comparison to experimental data from 
heavy ion collisions \cite{Romatschke:2007mq} the value of the shear viscosity to entropy density ratio $\eta/s$
could be determined to be close to the conjectured strong
coupling limit of $\eta/s=1/4\pi$ obtained from the AdS/CFT correspondence 
\cite{Policastro:2001yc,Kovtun:2004de}. Since then much progress has
been made. In particular the inclusion of
event-by-event fluctuations in the initial state of viscous hydrodynamics
\cite{Schenke:2010rr,Bozek:2012fw,Karpenko:2015xea} has turned out to be of great importance.
For recent reviews on the status of
relativistic fluid dynamics for heavy ion collisions see
\cite{Heinz:2013th,Gale:2013da,deSouza:2015ena,Jeon:2015dfa}.

\subsection{Relevance of initial state fluctuations for data interpretation}
The role of fluctuations in the transverse geometry of the collision
system was realized when experimentally studying the elliptic flow in
central Cu+Cu collisions at RHIC \cite{Alver:2006wh}. The large
observed $v_2$ could only be explained when the shape of the overlap
region was calculated relative to an axis determined by the fluctuating participants.
This concept turned out to have far reaching consequences.
In particular it allowed for the explanation of the structure of two
particle correlations in their pseudo-rapidity $\eta_p$ and azimuthal
angular difference, namely the so called ridge.
Its structure around $\Delta \phi=\pi$ in central collisions is
due to the contribution of the odd harmonic $v_3$, which in the
absence of fluctuations would be zero
\cite{Mishra:2007tw,Alver:2010gr,Alver:2010dn}.

The combined analysis of all $v_n$ and their event-by-event
distributions \cite{Aad:2013xma} has allowed to constrain the 
initial state and its fluctuations as well as the shear viscosity of the medium. 
The event-by-event distributions when scaled by the mean $v_n$
are largely independent of the detailed transport
parameters of the medium \cite{Niemi:2012aj}, which makes them ideal
observables to constrain features of the initial state.  In fact,
currently there are only a few initial state models that describe
all $v_n$ distributions $(n=2,3,4)$ for all experimentally measured
centralities. Most prominently, those are the IP-Glasma model
\cite{Schenke:2012wb,Schenke:2012hg,Gale:2012rq}, that we will discuss
in more detail below, and the EKRT framework
\cite{Niemi:2015qia}. Both models include saturation effects and lead
to similar energy deposition in the transverse plane.
In \cite{Moreland:2014oya} it was shown that the relevant feature 
to describe the experimental data is that the initial entropy density 
is proportional to the product of thickness functions.

In Section \ref{sec:hydroAndData} we will describe in more detail the
relevance of the initial energy deposition and fluctuations for describing experimental data using the
IP-Glasma model coupled to fluid dynamic calculations.

\subsection{Fast thermalization: Is it necessary?}
Hydrodynamics is an expansion around the energy momentum tensor of an
ideal fluid\footnote{In recent years, it has been proposed to expand
  around a fluid whose energy-momentum tensor is not isotropic
  \cite{Martinez:2010sd,Martinez:2012tu,Florkowski:2013lza,Bazow:2013ifa,Strickland:2014pga,Tinti:2015xra,Nopoush:2015yga}.}
at rest. Since the baseline of this expansion is a fluid in local
thermal equilibrium, it is often assumed that near-equilibrium is a
prerequisite for hydrodynamics.

Data on flow observables indicate a very effective
transfer from spatial anisotropy to momentum anisotropy. However,
this transfer would be impaired by the strong dissipative effects
that occur during the rearrangement of the internal degrees of
freedom of an off-equilibrium system. Moreover, for a
successful description of bulk observables in heavy ion collisions,
the hydrodynamical evolution should start very shortly after the
collision, at times $\tau\lesssim 1$~fm/c.

However, there is no direct evidence from data that the system
is indeed close to equilibrium. Although the above argument suggests
that a certain amount of pre-equilibration must have taken place
before hydrodynamics becomes a valid description, there are examples
(exactly solvable AdS/CFT models \cite{Heller:2011ju}, or
systems also studied in kinetic theory
\cite{Kurkela:2015qoa}) where the deviation of the energy-momentum
tensor from its ideal form is of order one, and hydrodynamics
nevertheless manages to track correctly the bulk evolution.

These examples suggest less stringent requirements: there should be a
range of time where the pre-hydrodynamical description and
hydrodynamics agree on the evolution of the stress tensor, even if it
is still off-equilibrium. But even this weaker condition is hard to
achieve in QCD. At leading order, QCD-based descriptions lead to an
increasing bulk anisotropy, while it decreases in hydrodynamics. As we
shall see later, higher order quantum fluctuations are essential in
the isotropization of the stress tensor.

\section{Initial state in $2+1$ dimensional classical Yang-Mills}
\subsection{QCD and Color Glass Condensate}\label{sec:MV}
Asymptotic freedom ensures that QCD perturbation theory can be used
for processes involving a hard momentum scale. However, its
applicability to the numerous softer particles is a priori
questionable.

\begin{figure}[htbp]
 \begin{minipage}{0.37\textwidth}
   \includegraphics[width=6.5cm]{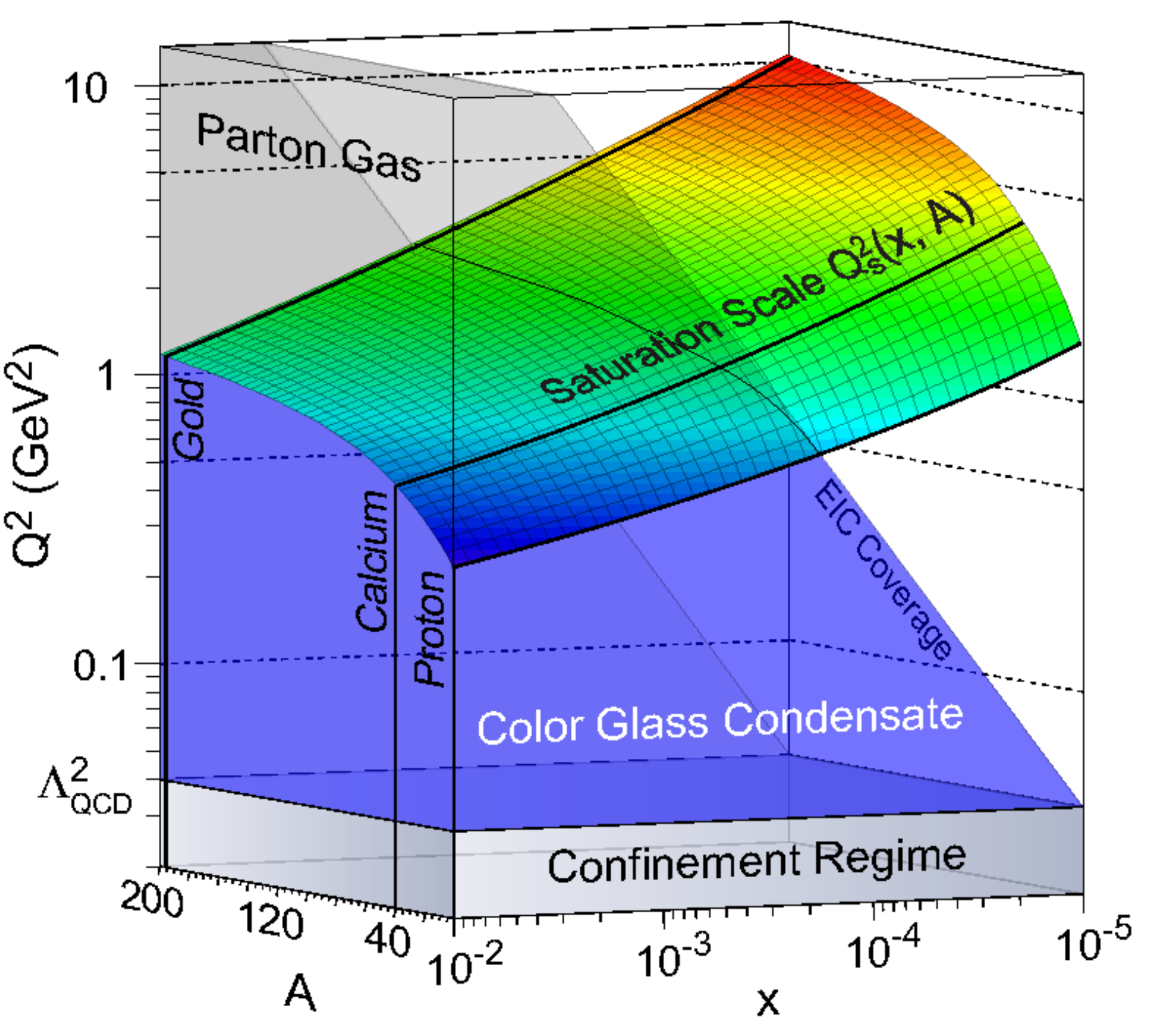}
  \end{minipage}
 \begin{minipage}{0.3\textwidth}
   \includegraphics[width=4.cm]{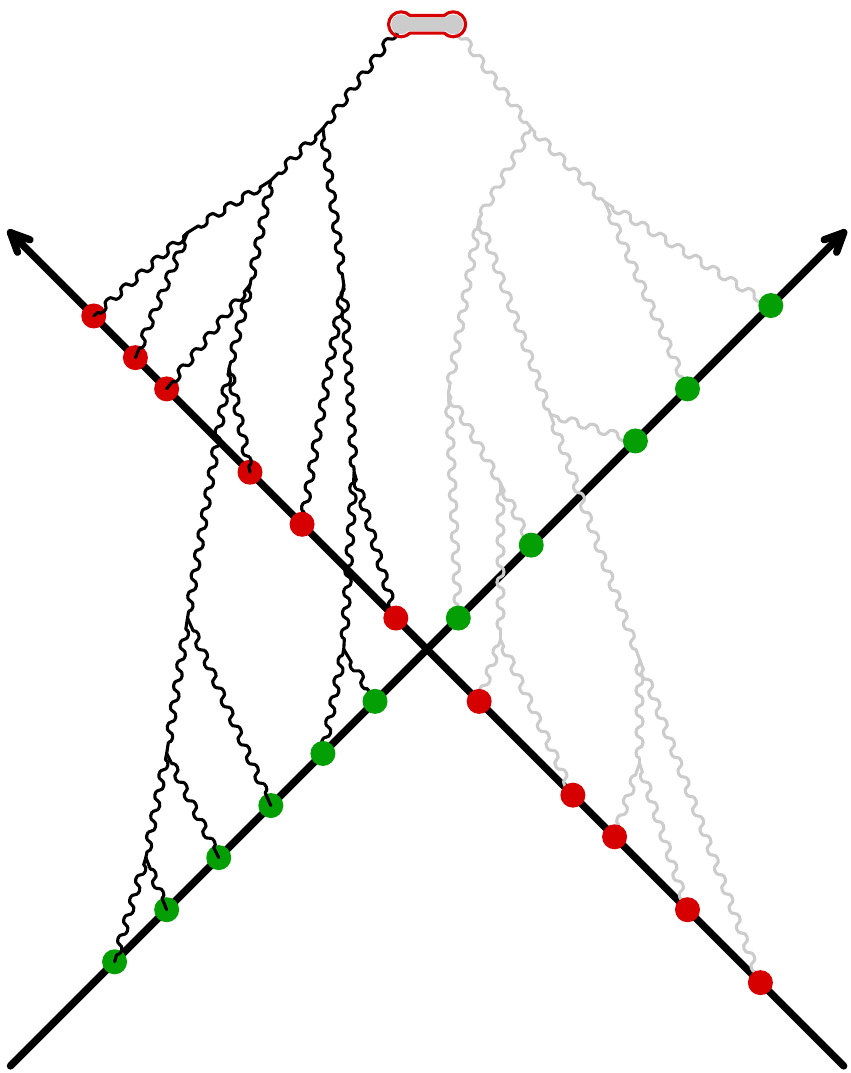}
  \end{minipage}
 \begin{minipage}{0.3\textwidth}
   \includegraphics[width=4.cm]{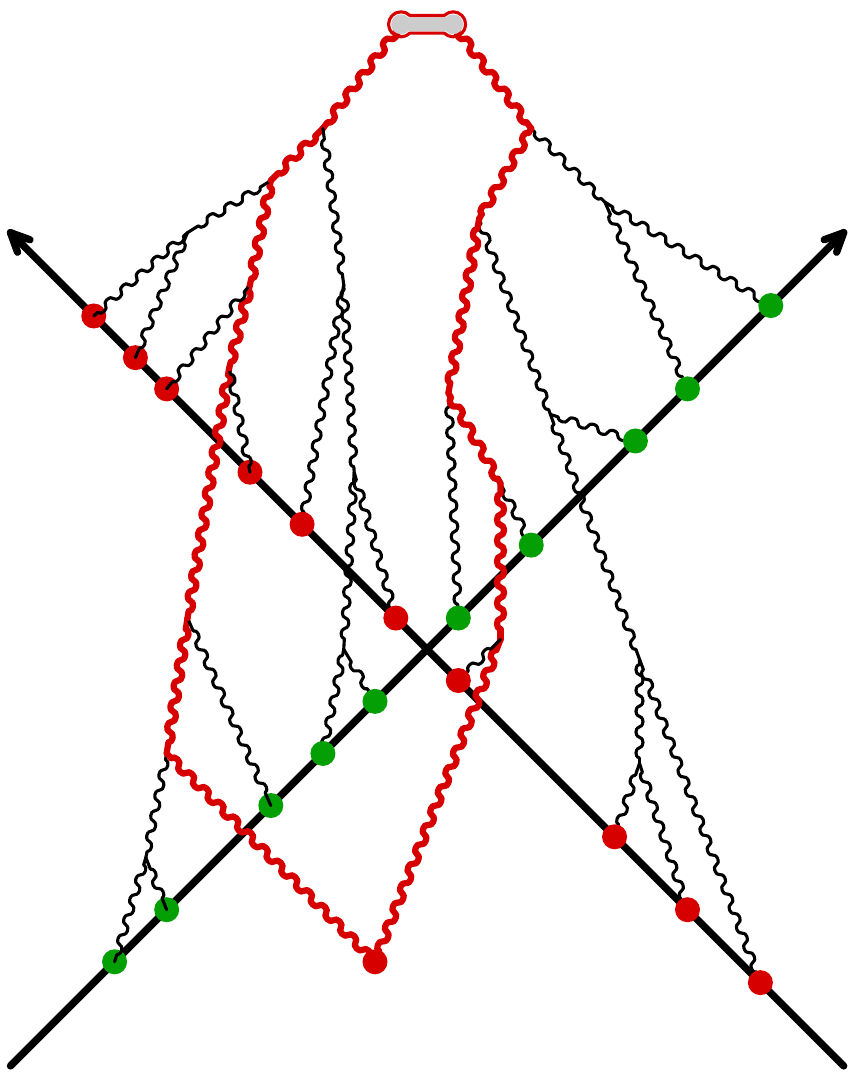}
  \end{minipage}
\caption{\label{fig:satdom}Left: Saturation domain depending on $x$,
  $Q$ and $A$. From \cite{Deshpande:2009zz}. Middle: Connected tree
  graphs that contribute to the single gluon spectrum at leading
  order (LO). The red and green dots denote the color sources of the two
  projectiles. Right: Same at NLO.}
\end{figure}

The gluon distribution in a hadron becomes large at small momentum
fraction $x$ and fixed transverse scale $Q^{-1}$. At fixed $Q$ and
decreasing $x$, gluons must eventually overlap in phase-space. When
their occupation number is comparable to $\alpha_s^{-1}$, gluon-gluon
interactions become important. In particular, gluon recombinations
tend to stabilize the occupation number, a phenomenon known as
\emph{gluon saturation}. New gluons can be produced only in the tail
of the distribution, which is not yet saturated. Consequently, the
typical gluon momentum increases with energy, leading to the
saturation momentum $Q_s(x)$ depicted in
Fig.\,\ref{fig:satdom}. Moreover, thanks to asymptotic freedom, the
corresponding value of the strong coupling constant decreases.

Gluons at the edge of saturation dominate scattering processes,
allowing a weak coupling treatment for the calculation of the bulk of
particle production. However, this system is non-perturbative, since a
large occupation number of order $\alpha_s^{-1}$ compensates the
smallness of the coupling. The \emph{Color Glass Condensate} (CGC) is
a QCD-based effective theory, designed to organize calculations in the
saturation regime \cite{Gelis:2010nm}. Its central idea is to use the
large gluon occupation number in order to organize the expansion
around classical solutions.

An observer in the center of momentum frame of a collision sees two
streams of color charges flowing from opposite directions, that can be
represented as two color currents $J_1^\mu$ and $J_2^\mu$. At high
energy, their dominant components are the $J^\pm$\footnote{We use
  light-cone coordinates: $x^\pm\equiv(x^0\pm x^3)/\sqrt{2}$.}
($J^{1,2}$ are inversely proportional to the collision energy, and
thus neglected).  Because of time dilation, the internal dynamics of
the projectiles appears totally frozen to the observer, and therefore
$J_1^\mu$ (resp. $J_2^\mu$) is independent of $x^+$ (resp. $x^-$):
\begin{equation}
J_1^{\mu a}(x)=\delta^{\mu+}\rho_{1a}(x^-,\x_\perp)
\quad,\qquad
J_2^{\mu a}(x)=\delta^{\mu-}\rho_{2a}(x^+,\x_\perp)\; ,
\end{equation}
where the functions $\rho_{1,2}$ represent the density of color
charges in the projectiles. These distributions reflect the
configuration of the color charges just before the collision and are
not known event-by-event, but one may develop a theory for its
statistical distribution $W[\rho]$. The McLerran-Venugopalan
\cite{McLerran:1994ni,McLerran:1994ka} model argues that in a nucleus
with many colored constituents this distribution should be
Gaussian. Moreover, thanks to confinement, this model neglects
correlations between partons located at different transverse
positions:
\begin{equation}
  W[\rho]\equiv\exp\Big\{-\int d^2\x_\perp \frac{\rho_a(x^-,\x_\perp)\rho_a(x^-,\x_\perp)}{2\mu^2(x^-,\x_\perp)}\Big\}\; ,
  \label{eq:MVdist}
\end{equation}
where $\mu^2(x^-,\x_\perp)$ characterizes the local density of color charges.

Closer to the observer's rapidity, degrees of freedom should not be
approximated by static sources, but treated as conventional quantum
fields. Thus, the CGC is a Yang-Mills theory coupled to an external
color current,
\begin{equation}
{\cal L}_{_{\rm CGC}}\equiv -\frac{1}{2}{\rm tr}\,\big(F^{\mu\nu}F_{\mu\nu}\big)
+ J_\mu A^\mu\; .
\end{equation}  
The coupling $J_\mu A^\mu$ is eikonal because the two types of degrees
of freedom have vastly different longitudinal momenta. In order to
avoid contributions from loop corrections that are already included in
the sources, CGC calculations beyond LO require the
introduction of cutoffs (one for each projectile) in longitudinal
momentum in order to properly separate sources from fields.

The order of magnitude of a connected graph ${\cal G}$ is
\begin{equation}
{\cal O}({\cal G})={g}^{-2} g^{n_{_E}} g^{2n_{_L}} (gJ)^{n_{_J}}\; ,
\end{equation}
where $n_{_E}$ counts the external gluons, $n_{_L}$ the independent
loops and $n_{_J}$ the sources $J_{1,2}^\mu$ in the graph. In the
saturated regime, $gJ\sim g^0$, and the order of magnitude depends
only on $n_{_E}$ and $n_{_L}$. Infinitely many graphs, differing in
the number of sources, contribute to a given order in $\alpha_s$:
despite a weak coupling, the CGC is non-perturbative.

Crucial simplifications occur for \emph{inclusive observables},
obtained as an average over all final states.  In particular, only
connected graphs contribute to such observables. The simplest
inclusive observable is the single gluon spectrum, obtained at leading
order by summing all the tree graphs shown in Fig.\,\ref{fig:satdom}.
These tree graphs can be expressed in terms of a solution to
the classical Yang-Mills equations,
\begin{equation}
  \big[{\cal D}_\mu,{\cal F}^{\mu\nu}\big]=J_1^\nu+J_2^\nu\;,\quad
  \lim_{x^0\to-\infty}{\cal A}^\mu(x)=0\; ,
\end{equation}
in the Fock-Schwinger gauge $x^+A^-+x^-A^+=0$.  Intuitively, the
absence of constraints on the gauge field at $x^0\to+\infty$ comes
from the fact that one sums over all final states.

From this classical solution, the gluon
spectrum at LO is given by
\begin{equation}
\left.
\frac{d{{\colora{N_1}}}}{dY d^2\vec\p_\perp}\right|_{_{\rm LO}}=
\frac{1}{16\pi^3}\int_{x,y}\; {\colorc e^{ip\cdot (x-y)}}\;
\square_x\square_y\;
\sum_\lambda \epsilon^\mu_\lambda \epsilon^\nu_\lambda\;\;
{\colord {\cal A}_\mu(x)}{\colord {\cal A}_\nu(y)}\; ,
\end{equation}
where $\square_x$ is the D'Alembertian operator and the
$\epsilon_\lambda^\mu$ are polarization vectors.  Likewise, the
multi-gluon spectra read
\begin{equation}
\left.
\frac{d{{\colora{N_n}}}}{d^3\p_1\cdots d^3\p_n}\right|_{_{\rm LO}}=
\left.
\frac{d{{\colora{N_1}}}}{d^3\p_1}\right|_{_{\rm LO}}
\times\cdots\times
\left.
\frac{d{{\colora{N_1}}}}{d^3\p_n}\right|_{_{\rm LO}}\; .
\label{eq:Nn}
\end{equation}
Initial conditions for hydrodynamical models require the 
energy-momentum tensor, whose expression in terms of
the classical chromo-electric and chromo-magnetic fields ${\bs E}^i$
and ${\bs B}^i$ read
\begin{eqnarray}
&&
T^{00}_{_{\rm LO}}
=
\frac{1}{2}\big[{{\bs E}^2+{\bs B}^2}\big]
\qquad
T^{0i}_{_{\rm LO}}=\big[{\bs E}\times {\bs B}\big]^i
\\
&&
T^{ij}_{_{\rm LO}}
=
\frac{\delta^{ij}}{2}\big[{{\bs E}^2+{\bs B}^2}\big]
-\big[{\bs E}^i{\bs E}^j+{\bs B}^i{\bs B}^j\big]\; .
\label{eq:Tclass}
\end{eqnarray}

\subsection{Practical implementation}
\label{sec:practical}
\begin{figure}[htbp]
\begin{center}
\resizebox*{0.9 \textwidth}{!}{\includegraphics{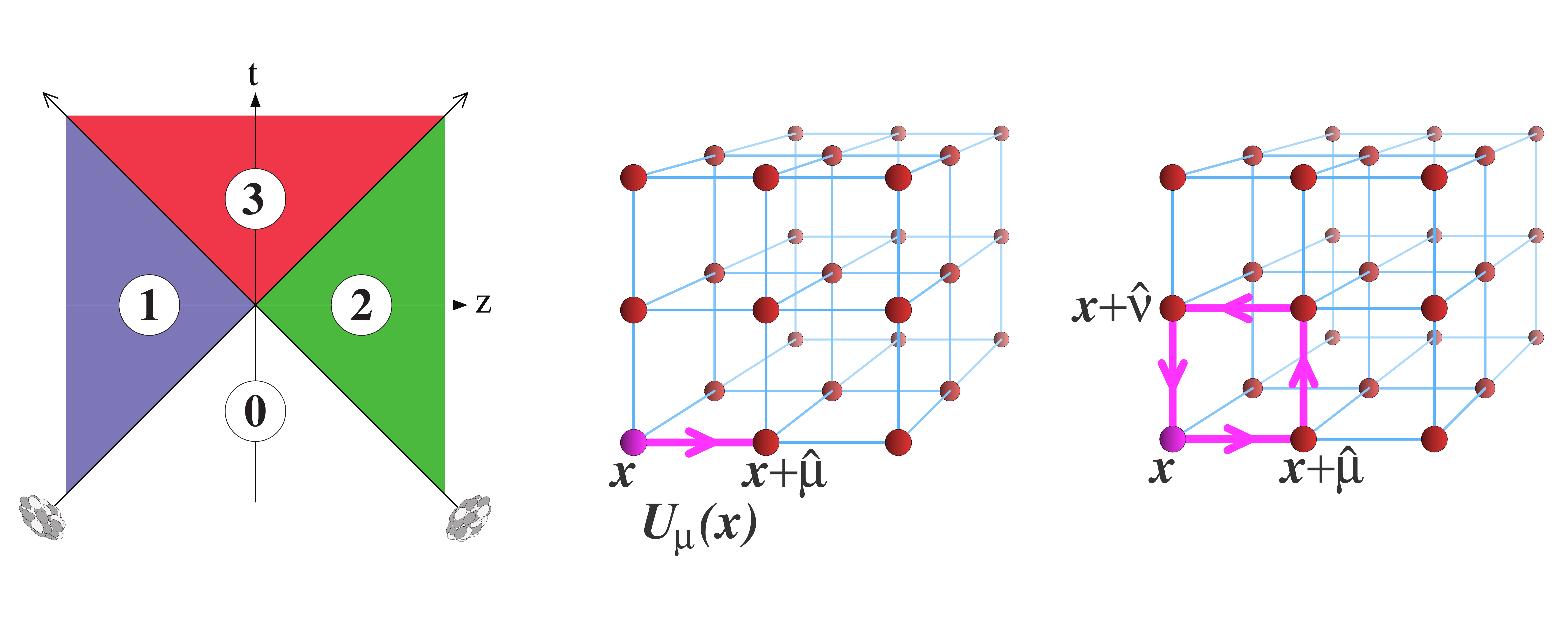}}
\end{center}
\caption{\label{fig:latt1}Left: Space-time structure of the classical
  gauge field ${\cal A}^\mu$. Middle and right: Link variable and
  plaquette on a 3-dimensional lattice.}
\end{figure}
The non-linear Yang-Mills equations cannot be solved analytically in
general. For numerical approaches, one should have in mind the
following:
\begin{enumerate}
\item~Since collisions at high energy are almost invariant under
  longitudinal boosts, it is natural to map the forward light-cone
  with proper-time ($\tau\equiv\sqrt{2x^+ x^-}$) and rapidity
  ($\eta\equiv\frac{1}{2}\log(x^+/x^-)$). The Yang-Mills equations do
  not depend on rapidity, and become 1+2 dimensional if their initial
  condition is itself independent of rapidity (which is the case in
  the CGC at LO).
\item~ The sources $\rho_{1,2}$ are singular on the light-cones
  $x^\pm=0$, that divide space-time in four regions shown in
  Fig.\,\ref{fig:latt1} (left). The gauge potential ${\cal A}^\mu$
  vanishes in region 0, and is known analytically in regions
  1,2~\cite{Kovchegov:1996ty}. In region 3, it is known analytically
  just above the light-cone, at $\tau=0^+$~\cite{Kovner:1995ja}~:
\begin{eqnarray}
&
A_0^i=\alpha_1^i+\alpha_2^i\quad,\;
&
E_0^i=0\quad,\;
\alpha_n^i=\frac{i}{g}U_n^\dagger \partial^i U_n\quad(n=1,2)\; ,
\nonumber\\
&
A_{0\eta}=0\quad,\;
&
E_0^\eta=i\frac{g}{2}[\alpha_1^i,\alpha_2^i]\;,
\label{eq:init-LO-tau0}
\end{eqnarray}
where the Wilson line $U_{1}(\x_\perp)$ reads
\begin{equation}
V_1(\x_\perp)
=
{\rm P}\,e^{ig\int dx^- \frac{1}{\nabla_\perp^2}\rho_1(x^-,\x_\perp)}
\label{eq:wilson-def}
\end{equation}
(and a similar expression for $V_2$.)
\end{enumerate}

Therefore, one needs to solve numerically 1+2-dim equations for
$\tau>0$~\cite{Krasnitz:1998ns,Krasnitz:1999wc,Krasnitz:2000gz,Krasnitz:2001qu,Krasnitz:2003jw,Lappi:2003bi,Lappi:2006hq,Krasnitz:2002ng,Krasnitz:2002mn,Lappi:2006xc},
with initial conditions (\ref{eq:init-LO-tau0}) at $\tau=0^+$.  In
practice, space is discretized on a lattice (see
Fig.\,\ref{fig:latt1}), while time remains a continuously varying
variable. One uses Wilson's formulation, where the gauge potentials
$A^\mu$ are replaced by link variables (see Fig.~\ref{fig:latt1}),
i.e. Wilson lines that span one elementary edge of the lattice
\begin{equation}
  {\colorc U_i(x)} \equiv {\rm P}\,\exp i\,{\colord g}\int_x^{x+\hat{\imath}}ds\; A^i(s)\; .
\end{equation}
In contrast, the electrical fields $E^i$ should be assigned to the
nodes of the lattice, and the discretized
Hamiltonian reads
\begin{eqnarray}
      {\cal H}&=&
\sum_{\vec\x;i}\frac{E^i(\x) E^i(\x)}{2}\nonumber\\
&&-\frac{6}{{\colord g^2}}\sum_{\vec\x;ij}1-\frac{1}{3}{\rm Re\ Tr}\,(\underbrace{U_i(x)U_j(x+\hat{\imath})U^\dagger_i(x+\hat{\jmath})U^\dagger_j(x)}_{\mbox{\scriptsize plaquette at the point $\vec\x$ in the $ij$ plane}})\; .
\end{eqnarray}
The corresponding Hamilton equations form a large but finite set of
ordinary differential equations, that can be solved e.g. by
the leapfrog algorithm.

Shortly after the collision (at $\tau\ll Q_s^{-1}$), the classical
chromo-electric and chromo-magnetic fields are aligned with the collision
axis~\cite{Lappi:2006fp}.  The expectation value of transverse Wilson
loops~\cite{Dumitru:2013koh,Dumitru:2014nka},
\begin{equation}
W\equiv \left<{\rm P}\,\exp ig\int_{\gamma}dx^i {\cal A}^i\right>\; ,
\end{equation}
that measures the magnetic flux through the loop, provides information
on the transverse correlation length of these fields.  For large loops
of area larger than $Q_s^{-2}$, $W$ decreases approximately as
$\exp(-\#\times\mbox{Area})$, indicating a
decorrelation on transverse distances larger than $Q_s^{-1}$.

\subsection{IP-Glasma framework} \label{sec:ipglasma}
As discussed above, 
Yang-Mills equations for the boost
invariant system
have to be solved numerically,
which has been done for homogeneous nuclei and $N_c=2$ in
\cite{Krasnitz:1998ns} and for $N_c=3$ in
\cite{Krasnitz:2001qu}. Finite size nuclei were studied in
\cite{Krasnitz:2002mn,Lappi:2003bi}. In particular, in
\cite{Krasnitz:2002mn} nucleons were sampled from a Woods-Saxon
distribution and the color charge density of the nucleus was taken to
be proportional to the sum of the thickness functions of all
nucleons. Coulomb tails were avoided by implementing a color
neutrality condition on length scales given by the inverse of
$\Lambda_{\rm QCD}$.

 \begin{figure}[h]
   \includegraphics[width=1.0\textwidth]{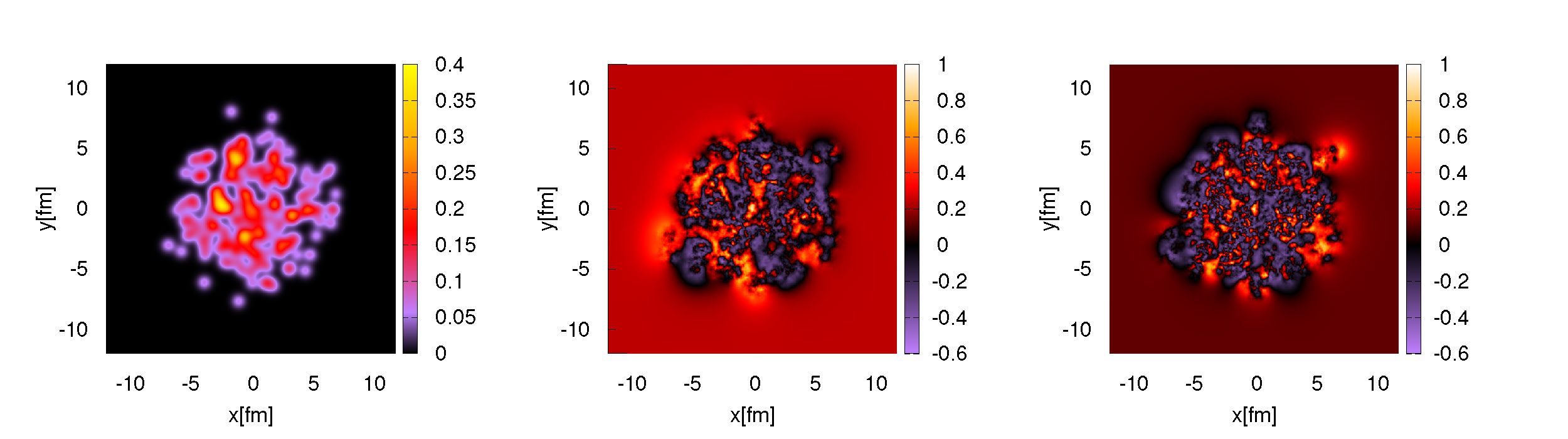}
   \caption{Left: The incoming color charge density $g^2\mu^2$ for a
     gold nucleus at $\sqrt{s}=200\,{\rm GeV}$. Middle: The correlator
     $(1/N_c) {\rm Re} [{\rm Tr}(V^\dag(0,0) V(x,y))]$ showing the
     degree of correlations in the gluon fields for a gold ion at
     $\sqrt{s}=200\,{\rm GeV}$ at an $x$ value that will contribute to
     gluon production at mid-rapidity. Right: Same as the middle figure
     but for $\sqrt{s}=5\,{\rm TeV}$. \label{fig:gmu} }
 \end{figure}

The IP-Glasma model \cite{Schenke:2012wb,Schenke:2012hg} is very similar to the framework introduced
in \cite{Krasnitz:2002mn}. The main differences are the use of the
IP-Sat model \cite{Bartels:2002cj,Kowalski:2003hm} to constrain the
$x$ and transverse position dependence of the color charge density
using data from deeply inelastic scattering experiments, and the way
one deals with the infrared tails. We now give a brief description of
the various steps involved in computing the fluctuating initial state
in the IP-Glasma model.

\begin{enumerate}
\item Nucleon positions $\x_\perp^i$ ($i=1\dots A$) in the
  transverse plane of two nuclei are sampled from Woods-Saxon
  distributions with parameters adjusted to the nucleus of interest.
\item The sum of the Gaussian nucleon thickness functions $T_p$
is computed. It enters the IP-Sat expression for the dipole cross section in deeply
inelastic scattering \cite{Kowalski:2007rw}
\begin{align}
&\frac{1}{2}\frac{\ud \sigma^{\textrm{A}}_{\textrm{dip}}}{\ud^2 \xt}(\rt,\xt,x)=\mathcal{N}_A(\rt,\xt,x)\notag\\
&~~~~=\left[1-e^{-\frac{\pi^2}{2N_{c}}\r_\perp^2\alpha_{s}(Q^{2}) xg(x,Q^{2})\sum_{i=1}^A T_p(\xt-\x_\perp^i)}\right]\,.
\label{eq:nuc-dipole}
\end{align}
$\mathcal{N}_A$ is the scattering amplitude of the nucleus, $Q$ is the
momentum scale related to the dipole size $\rt$,
$Q^2=4/\r_\perp^2+Q_0^2$, with $Q_0$ fixed by the HERA
inclusive data. In the first studies
\cite{Schenke:2012wb,Schenke:2012hg}, parameters were taken from the
fit in Ref.~\cite{Kowalski:2006hc}, later \cite{Schenke:2013dpa}
parameters from fits to high precision combined data from the H1 and
ZEUS collaborations~\cite{Rezaeian:2012ji} were used.  The gluon
distribution $xg(x,Q^{2})$ is parameterized at the initial scale
$Q_0^2$ as $xg(x,Q^{2}_{0})=A_{g}x^{-\lambda_{g}}(1-x)^{5.6}$ and then
evolved up to the scale $Q^2$ using LO
DGLAP-evolution. Like $Q_0$, $A_g$ and $\lambda_g$ are constrained by
the fit to HERA data.
\item Using the definition of the nuclear saturation scale $Q_s$ as
  the inverse value of $r=\sqrt{\r_\perp^2}$ for which
  $\mathcal{N}_A = 1 - e^{-1/2}$, $Q_s(\xt, x)$ is extracted. Using
  their proportionality, the color charge squared per unit area
  $g^2\mu^2(\xt, x)$ is obtained from $Q_s(\xt, x)$. The
  proportionality factor depends on the details of the calculation
  \cite{Lappi:2007ku} and in the IP-Glasma calculations it is allowed
  to be varied in order to reproduce the overall normalization of
  produced particles \cite{Schenke:2012hg}.
  A typical distribution of $g^2\mu^2(\xt)$ is shown in
  Fig.\,\ref{fig:gmu} (left).
\item For a given $x$, which depends on the energy of the collision
  and the rapidity of interest,
  color charges $\rho^a(\xt)$ are sampled from the distribution (\ref{eq:MVdist}).
\item Assuming a finite width of the nucleus, the discretized version of the Wilson line (\ref{eq:wilson-def}) is given by \cite{Lappi:2007ku}
\begin{equation}
  V(\xt) = \prod_{k=1}^{N_y}\exp\left(-ig\frac{\rho_k(\xt)}{\boldsymbol{\nabla}_\perp^2+m^2}\right)\,,
\end{equation}
where $m\sim\Lambda_{\rm QCD}$. The correlator of these Wilson lines $1/N_c {\rm Re} [{\rm
    Tr}(V^\dag(0) V(\xt))]$ for two nuclei at different energies is shown in
Fig.\,\ref{fig:gmu} (middle and right). The scale of the fluctuations
of this quantity is $1/Q_s(\xt)$, which is smaller for the
higher energy case (right).
\item For each nucleus an SU($N_c$) matrix $V_j$ is assigned at each
  lattice site $j$.  They define a pure gauge configuration with the
  link variables
\begin{equation}\label{eq:links}
  U^{i}_{j} = V_{j}V^\dag_{j+\hat{e_i}}\,,
\end{equation}
where $+\hat{e_i}$ indicates a shift from $j$ by one lattice site in the $i=1,2$ direction. 
\item The Wilson lines in the future light-cone
  $U^{i}_j$ are determined from those of the two nuclei ($A$ and
  $B$) by solving
\begin{align}
  &{\rm tr} \left\{ t^a \left[\left(U^{i}_{(A)}+U^{i}_{(B)}\right)(1+U^{i\dag})
   -(1+U^{i})\left(U^{i\dag}_{(A)}+U^{i\dag}_{(B)}\right)\right]\right\}=0\label{eq:initU}
\end{align}
iteratively \cite{Krasnitz:1998ns}. Here $t^a$ are the generators of $SU(3)$ in the fundamental representation.
\item The lattice expression for the longitudinal electric field can
  then be obtained from the solutions $U^{i}$ and $U^{i}_{(A,B)}$
  \cite{Krasnitz:1998ns}.
\item Given these initial conditions, the source free Yang-Mills
  equations are solved forward in time (see Section \ref{sec:practical}).
\end{enumerate}

\begin{figure}[h]
  \centering{\includegraphics[width=10cm]{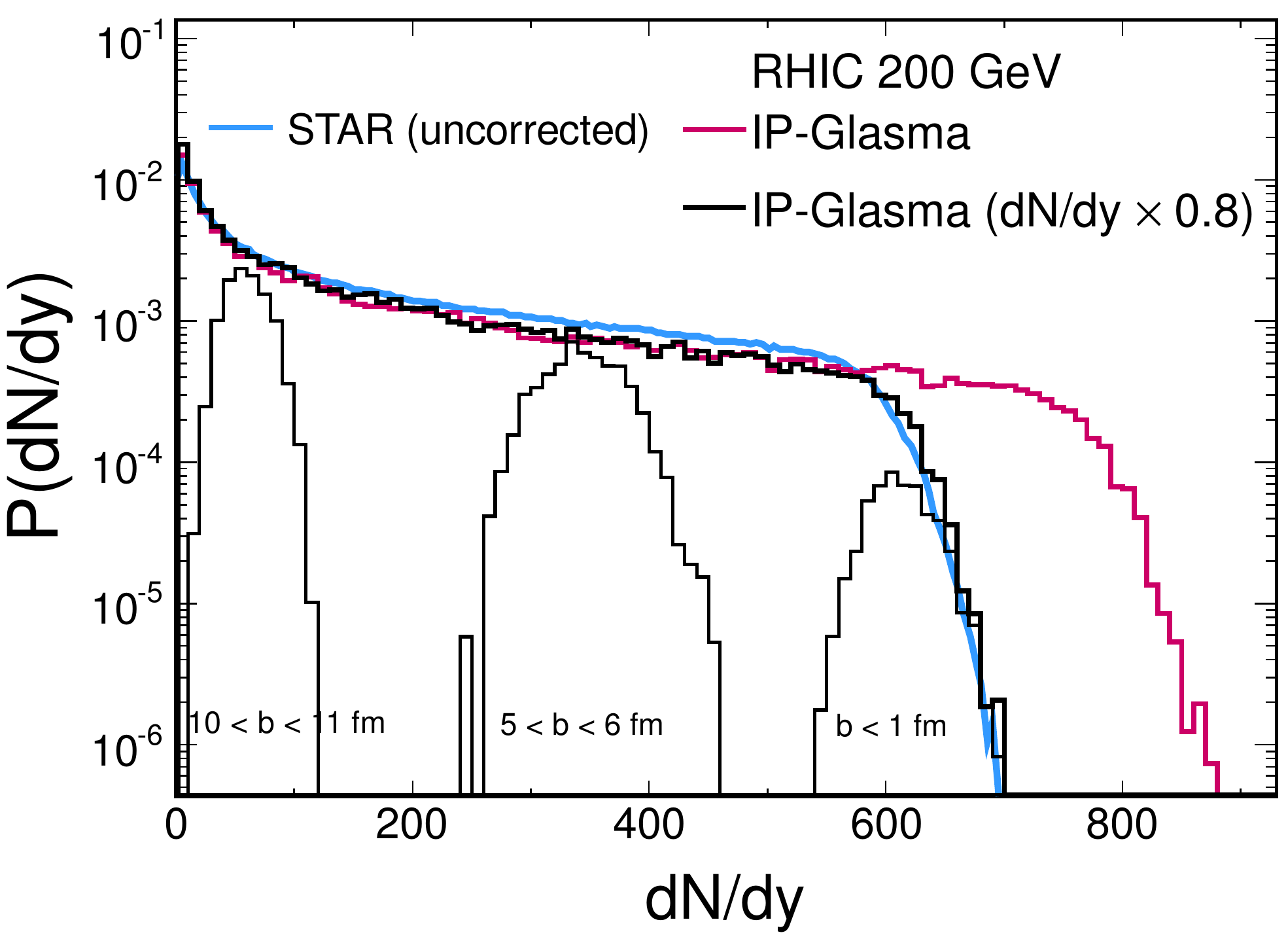}}
  \caption{Gluon multiplicity distribution. Shown are also distributions for
    limited ranges of impact parameter $b$. Experimental data from STAR
    \cite{Abelev:2008ez}. Figure from
    \cite{Schenke:2012hg}. \label{fig:multdist_rhic_paper} }
\end{figure}

From the boost invariant gauge field configurations at finite times
one can determine the gluon multiplicity and the energy momentum
tensor as a function of transverse position. 
Employing the proper impact parameter distribution
\cite{Schenke:2012hg} obtained from the Glauber model, the gluon
multiplicity distribution can be computed in transverse Coulomb gauge $\partial_i A^i =0$, $i=1,2$,
and compared to the measured charged hadron multiplicity distribution. The result for RHIC energies
comparing to uncorrected data from the STAR collaboration is shown in
Fig.\,\ref{fig:multdist_rhic_paper}. It was shown \cite{Gelis:2009wh}
that gluons produced from the Glasma naturally follow negative
binomial distributions. 
This can be seen from the distributions for small impact parameter ranges in 
Fig.\,\ref{fig:multdist_rhic_paper}. It was
demonstrated in \cite{Schenke:2012wb} that these distributions are
indeed fit best by negative binomial distributions.

\subsection{Connection to hydrodynamics and comparison to data} \label{sec:hydroAndData}
We now discuss how the classical color fields from the IP-Glasma
calculation are translated into an initial condition for
hydrodynamics, and show comparisons with
experimental data for several observables.

The main equations of hydrodynamics are energy and momentum
conservation $\partial_\mu T^{\mu\nu}=0$ along with an equation of state, where $T^{\mu\nu}$ is the
energy momentum tensor.
They need to be supplemented by an initial
condition for $T^{\mu\nu}$ at some early time, often
called the ``thermalization time''. The IP-Glasma model provides the gauge field
$T^{\mu\nu}$ with one caveat: the gluon field configurations of the
IP-Glasma model are not in local equilibrium. In fact, when switching from
IP-Glasma dynamics to hydrodynamics at a time of $0.2\,{\rm
  fm}/c$, the longitudinal pressure of the fields is approximately
zero, and negative at earlier times. We will discuss the
possibility to reach equilibrium in the Yang-Mills system when
including quantum corrections and considering fully three dimensional
dynamics below.  Lacking a mechanism for equilibration in the 2+1
dimensional LO theory, one way to provide an initial
condition for the hydrodynamic equations is to neglect the
non-equilibrium components of $T^{\mu\nu}$ and extract the energy
density and initial flow velocities by solving the identity $u_\mu
T^{\mu\nu}=\varepsilon u^{\nu}$ \cite{Gale:2012rq}. There will be discontinuities in other components of
$T^{\mu\nu}$ as one switches from the anisotropic field energy momentum
tensor to the equilibrium one, which contains the isotropic pressure
obtained from the equation of state. However, this should be a good
first approximation for matching to fluid dynamics. 

Here we will review results for observables that are
sensitive to the fluctuations in the initial state of the
collision. We will demonstrate that the IP-Glasma model produces fluctuating initial
geometries that are consistent with the flow harmonics $v_n$ and their
event-by-event fluctuations measured at the LHC.

The $v_n$ are the coefficients in a Fourier expansion of the azimuthal
charged hadron distribution, which is obtained after evolving the hydrodynamic
medium and performing a ``freeze-out'' wherever the system reaches a given minimal temperature
or energy density. Every cell, which reaches that
threshold, will act like a black body radiator of thermally
distributed particles \cite{Cooper:1974mv}.  Resonances then decay according to the
experimentally observed branching ratios, resulting in the final
particle distributions.

\subsubsection{Flow harmonics $v_n$}
Fluid dynamics translates initial geometries into momentum
anisotropies, which are quantified by the $v_n$. Thus, apart from the
right transport properties, a theoretical description must contain the
correct initial geometry including event-by-event fluctuations.
Fluctuations have a significant effect even on the average values of
the $v_n$, most notably for odd $n$: Without fluctuating initial
conditions the odd harmonics would be zero by symmetry.

\begin{figure}[ht]
 \begin{minipage}{0.49\textwidth}
  \includegraphics[width=\textwidth]{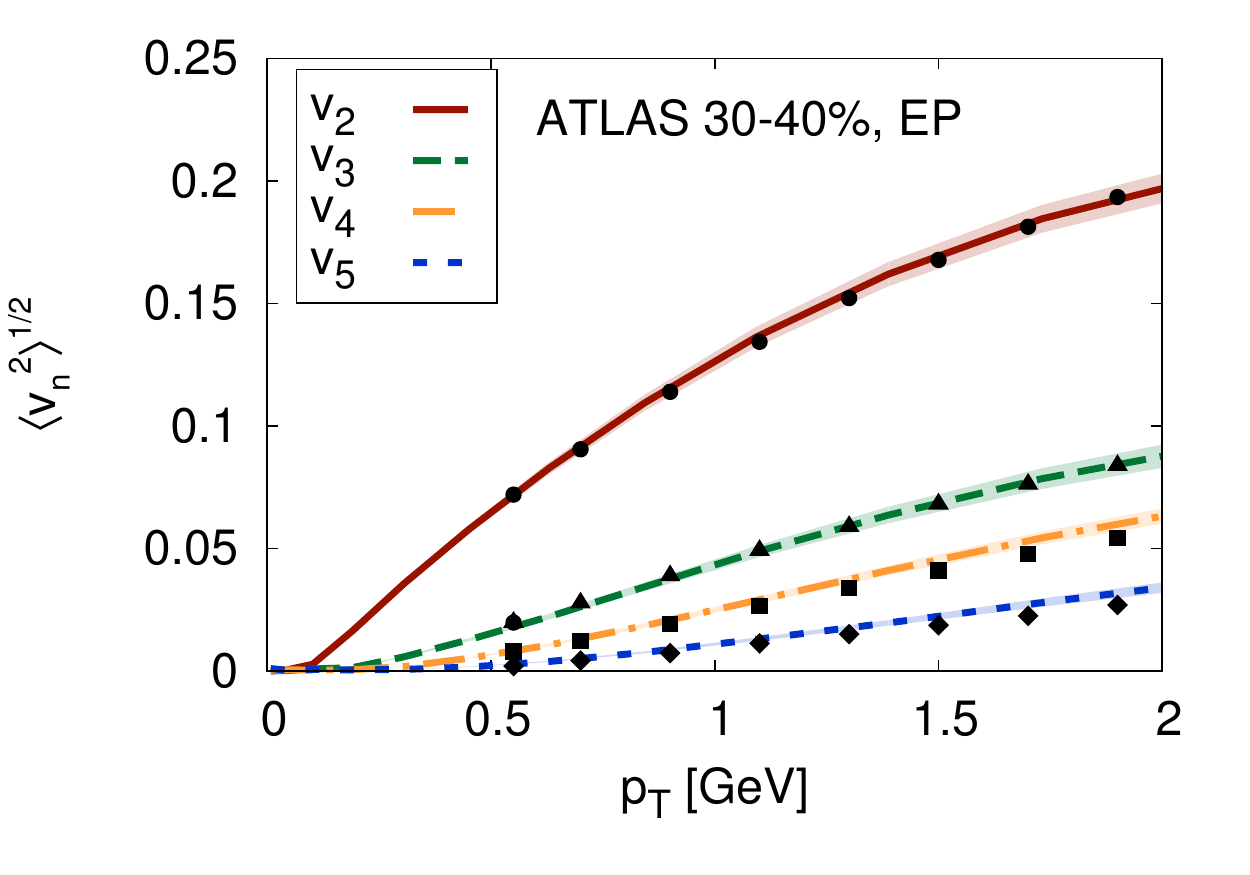}
  \end{minipage}
 \begin{minipage}{0.49\textwidth}
   \includegraphics[width=\textwidth]{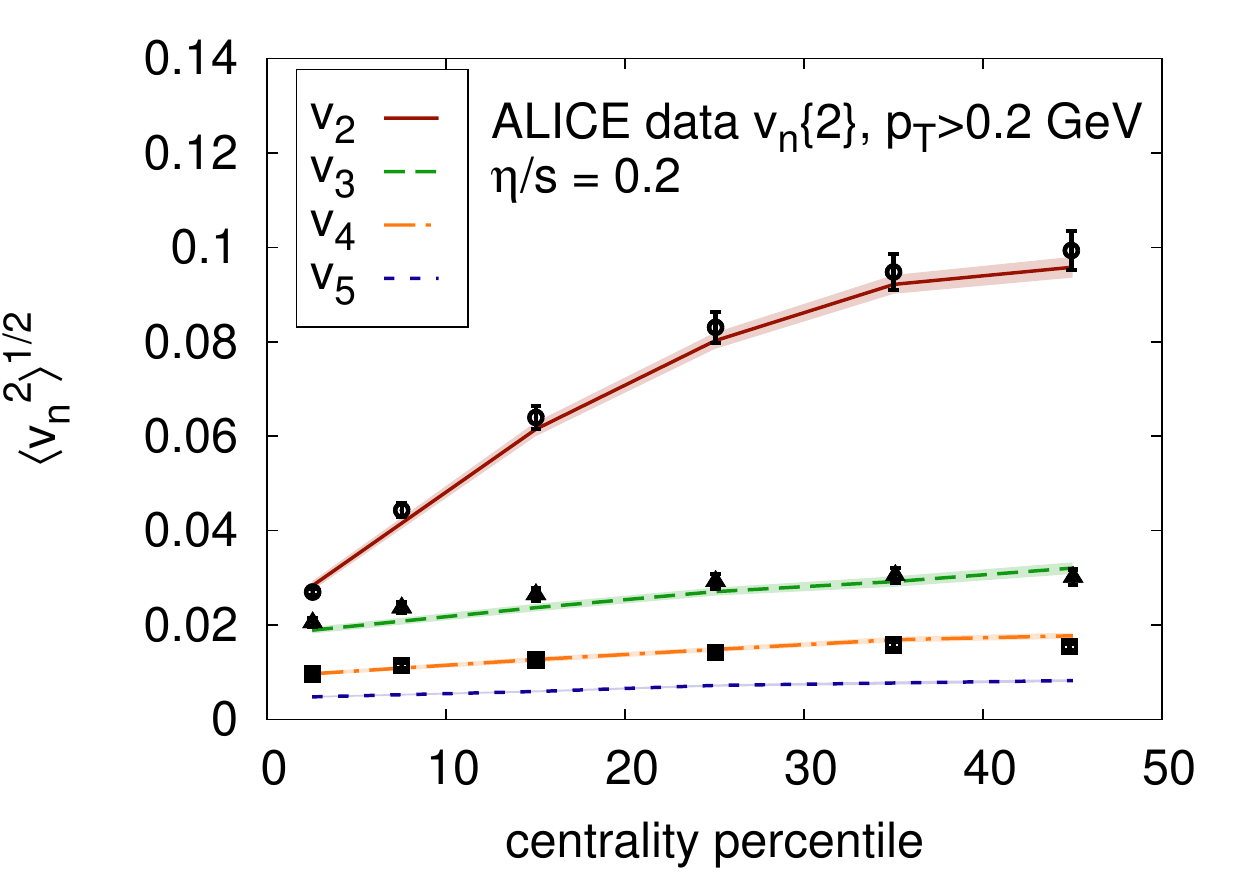}
  \end{minipage}
   \caption{ Left: Transverse momentum dependent rms $v_n$ compared to
     ATLAS data \cite{ATLAS:2012at}. Right: Centrality dependent rms
     $v_n$ compared to ALICE data \cite{ALICE:2011ab}. Figure adapted
     from \cite{Gale:2012rq}. \label{fig:vn30-40-LHC} }
\end{figure}

The IP-Glasma model in combination with viscous fluid dynamics
described above has been shown to lead to an exceptionally good
description of all flow harmonics, both as functions of transverse
momentum and collision centrality \cite{Gale:2012rq}. This is a
non-trivial result because other initial state models could not
describe $v_2$ and $v_3$ simultaneously \cite{Qiu:2011hf}. In
Fig.\,\ref{fig:vn30-40-LHC} we show $v_n$ ($n=2,\dots, 5$) as functions of $p_\perp$ and
centrality for Pb+Pb collisions at LHC with a center of mass energy of
$\sqrt{s}=2.76\,{\rm TeV}$.

\subsubsection{$v_n$ distributions}
An observable that is particularly sensitive to the initial state
fluctuations but almost insensitive to the transport properties of the
medium is the event-by-event distribution of harmonic flow
coefficients \cite{Aad:2013xma,Niemi:2012aj,Gale:2012rq}.  As
discussed in the introduction, several initial state models could be
excluded by comparing their predictions for the event-by-event
elliptic flow distributions in different centrality classes with
experimental data \cite{Niemi:2012aj}.

\begin{figure}[ht]
   \includegraphics[width=1 \textwidth]{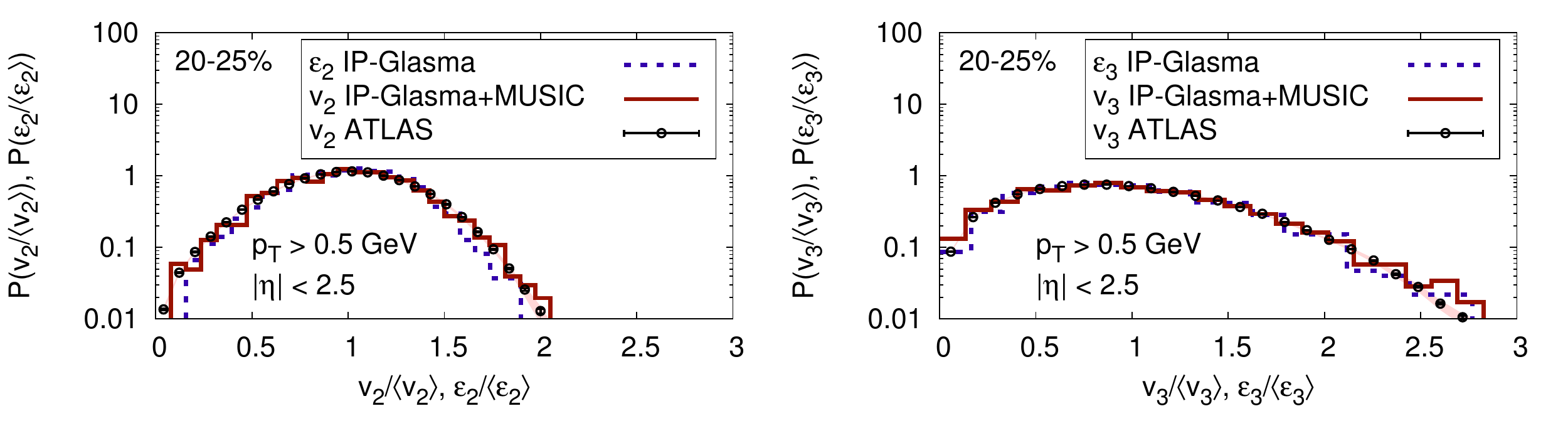}
   \caption{ Event-by-event distributions of elliptic (left) and
     triangular (right) flow coefficients in $\sqrt{s}=2.76\,{\rm TeV}$ collisions at LHC, scaled by the mean values and compared
     to ATLAS data \cite{Aad:2013xma}. Also shown are the scaled
     initial state eccentricity
     distributions. \label{fig:vnenDist-QM-20-25} }
 \end{figure}

Calculations of scaled event-by-event $v_n$ distributions using the
IP-Glas\-ma model combined with hydrodynamic evolution are in
exceptional agreement with experimental data
\cite{Gale:2012rq,Schenke:2014zha}. This is demonstrated for 20-25\%
central events in Fig.\,\ref{fig:vnenDist-QM-20-25}. For not too
peripheral events or too large harmonic number $n$ the initial
eccentricity distributions already yield a very good description of
the experimental data.
These results indicate that the initial state fluctuations used in the
IP-Glasma model describe the experimental reality very well.

\subsubsection{Shape-multiplicity correlations in U+U collisions}
Another observable that is sensitive to fluctuations and the mechanism
of particle production and able to exclude initial state models is the
correlation between multiplicity and elliptic flow in ultra-central
collisions of deformed nuclei such as
uranium U \cite{Adamczyk:2015obl}. A Glauber model in which the
multiplicity has a significant contribution proportional to the number
of binary collisions $N_{\rm bin}$ predicts a strong anti-correlation
between $v_2$ and the multiplicity in ultra-central U+U
collisions. This is because $N_{\rm bin}$ is much larger in the case
that the longer axes of both nuclei are aligned with the beam line,
than if the shorter axes are.

\begin{figure}[h]
   \centering{\includegraphics[width=0.7 \textwidth]{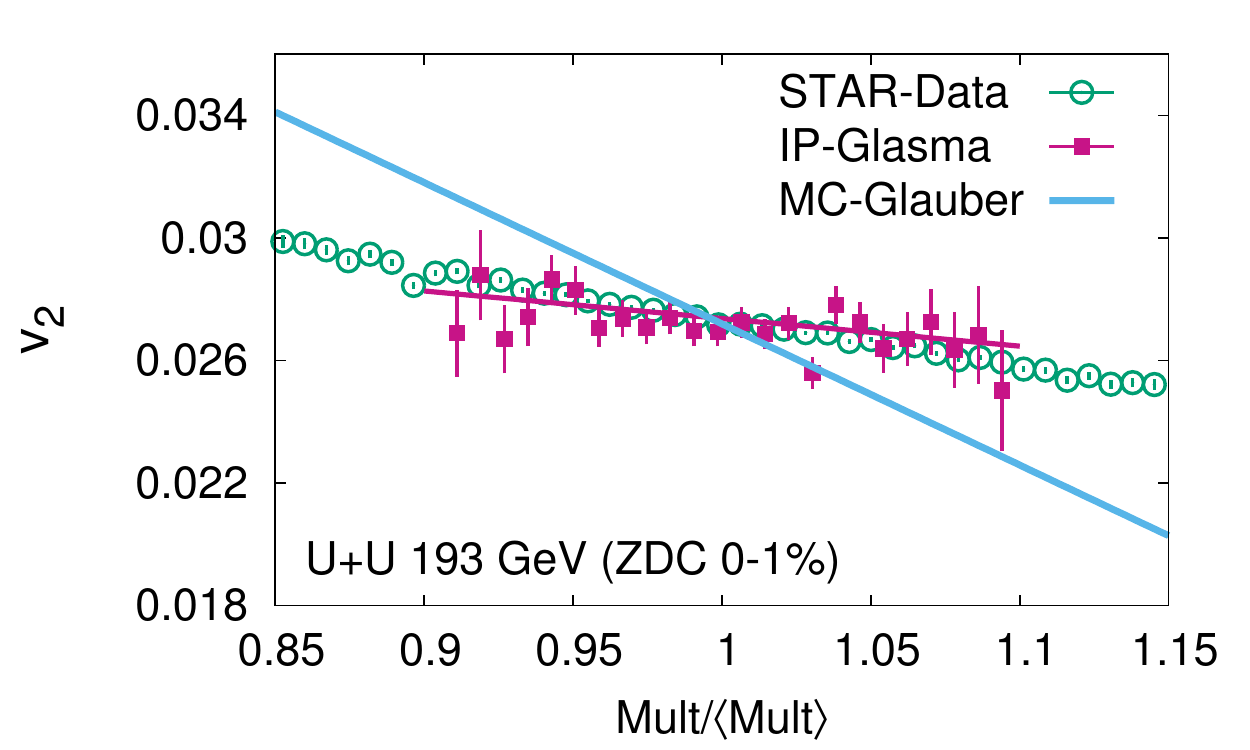}}
   \caption{ Correlation between $v_2$ ($\varepsilon_2 \langle
     v_2\rangle/\langle \varepsilon_2\rangle$ in the theoretical
     curves) and the scaled multiplicity in 0-1\% central U+U events
     at $\sqrt{s}=193\,{\rm GeV}$. The magenta line is a fit to the
     IP-Glasma result. Experimental data from STAR
     \cite{Adamczyk:2015obl}.
     \label{fig:ecc_zdc} }
\end{figure}

 In contrast, along with a simple constituent quark model, the
 IP-Glasma model predicts a weaker anti-correlation of the initial
 geometry with the multiplicity \cite{Schenke:2014tga}, which is very
 close to the experimental data as demonstrated in
 Fig.\,\ref{fig:ecc_zdc}.

In the IP-Glasma model the multiplicity is proportional to $Q_s^2 S_\perp/\alpha_s(Q_2)$
where $S_\perp$ is the transverse size of the overlap region. For
tip-tip collisions, which have the smallest $v_2$, the increase in
$Q_s^2$ is balanced by a decrease in $S_\perp$. $\alpha_s$ decreases
only logarithmically with increasing $Q_s^2$ leading to a mildly
increased multiplicity in tip-tip collisions compared to body-body
collisions (which have the largest $v_2$ due to the prolate shape of
the U nucleus).  The data indicate that effects from sub-nucleonic
structure and coherence, as included in the IP-Glasma model, are
present and important \cite{Adamczyk:2015obl}.

\section{Quantum fluctuations}\label{sec:fluct}
\subsection{General considerations}
The CGC at LO already contains fluctuations of the positions of the
nucleons inside a nucleus, and of the distribution of the color
charges inside a nucleon. But once the nucleon positions and the color
distribution in each nucleon has been chosen, the outcome is
deterministic.

At next-to-leading order (NLO), new fluctuations --of quantum origin--
appear and the fields are no longer determined deterministically from
the color sources. These quantum fluctuations can a priori be of two
kinds:
\begin{itemize}
  \item[{\bf i.}] {\bf Initial state fluctuations}. Because of the
    uncertainty principle, the fields and their conjugate momenta
    cannot be known with arbitrary accuracy. Therefore, a
    \emph{quantum} initial state must have fluctuations of the initial
    values of the fields, in contrast with the CGC at LO. The minimal
    variance of these fluctuations is controlled by Planck's constant
    $\hbar$.
  \item [{\bf ii.}] {\bf Quantum ``jumps'' in the time evolution}. In
    contrast with the LO where Hamilton equations provide unique time
    derivatives of the fields from their current values, quantum
    fluctuations also make the evolution non deterministic.
\end{itemize}

However, NLO corrections (one-loop) are rather special because they
only contain quantum fluctuations from the initial state (beyond NLO,
fluctuations are both in the initial state and in the evolution). In a
quantum system whose classical phase-space is described by coordinates
$\X$ and by momenta $\P$, the density matrix $\rho_t$ evolves
according to
\begin{equation}
  \partial_t\rho_t=i\big[H,\rho_t]\; ,
  \label{eq:vonN}
\end{equation}
where $H$ is the Hamiltonian operator. A completely equivalent
representation  is obtained by introducing the
\emph{Wigner representation},
\begin{eqnarray}
  {\cal W}_t(\X,\P)&\equiv&
  \int d\s\; e^{\tfrac{i}{\hbar}\P\cdot\s}\; \big<\X-\tfrac{\s}{2}\big|\rho_t\big|\X+\tfrac{\s}{2}\big>
  \nonumber\\
  {\cal H}(\X,\P)&\equiv&\int d\s\; e^{\tfrac{i}{\hbar}\P\cdot\s}\; \big<\X-\tfrac{\s}{2}\big|H\big|\X+\tfrac{\s}{2}\big>\; .
\end{eqnarray}
Note that $\X$ and $\P$ are commuting classical coordinates, not
quantum operators.  ${\cal W}_t$ is called the Wigner distribution,
and ${\cal H}$ is the classical Hamiltonian. ${\cal W}_t$ evolves as
\begin{eqnarray}
  \partial_t{\cal W}_t(\X,\P)
  &=&\frac{2}{\hbar}
  {\cal H}(\X,\P)\;
  \sin\left(\frac{\hbar}{2}\left(
  \stackrel{\leftarrow}{\nabla}_\X
  \stackrel{\rightarrow}{\nabla}_\P
  -\stackrel{\leftarrow}{\nabla}_\P
  \stackrel{\rightarrow}{\nabla}_\X
  \right)\right)\;
        {\cal W}_t(\X,\P)
        \nonumber\\
        &\empile{\approx}\over{\hbar\to 0}&
        \underbrace{\big\{{\cal H},{\cal W}_t\big\}}_{\mbox{\scriptsize Poisson bracket}}+{\cal O}(\hbar^2)\; .
\end{eqnarray}
The first line is exact, and the second line shows the lowest order in
$\hbar$. At the order $\hbar^0$, we recover classical Hamiltonian
dynamics, in the form of the Liouville equation. Remarkably, the first
correction arises only at the $\hbar^2$ level: at the order $\hbar^1$,
the time evolution remains classical, and the only quantum effects
come from the initial state.

\subsection{One loop corrections}
Consider now the 1-loop corrections to an observable ${\cal
  O}(A,\partial A,\cdots)$, that depends on the gauge field operator
and its derivatives.  One must evaluate graphs such as the one shown
in Fig.\,\ref{fig:satdom}, that contain a loop embedded in the LO
classical field \cite{Gelis:2006yv,Gelis:2007kn}. When its
\emph{endpoints have a space-like separation}\footnote{For time-like
  separations, the propagator receives extra contributions that depend
  on the time ordering \cite{Epelbaum:2011pc}.}, the gluon propagator
in a background field reads~\cite{Gelis:2008rw}:
\begin{equation}
{\cal G}^{\mu a, \nu b}(x,y)\empile{=}\over{(x-y)^2\le
  0}
\sum_{\lambda,c}\int\frac{d^3\k}{(2\pi)^3 2|\k|} \; a^{\mu a}_{\k\lambda c}(x)
\; a^{\nu b*}_{\k\lambda c}(y)\; ,
\label{eq:prop-back}
\end{equation}
where $a_{\k\lambda c}^{\mu a}(x)$ is a small perturbation obeying the
\emph{linearized} Yang-Mills equations about the classical field
${\cal A}^\mu$,
\begin{align}
\left[{\cal D}_{\mu},\left[{\cal D}^{\mu},a^{\nu}\right]
-\left[{\cal D}^{\nu},a^{\mu}\right]\right]
-ig\left[{\cal F}^{\nu\mu},a_{\mu}\right]=\null& 0\;,&
\lim\limits_{t\to-\infty}a^{\mu a}_{\k\lambda c}(x)=\null&\epsilon^{\mu}_{\k\lambda}\delta^a_c\; e^{ik\cdot x}\;.
\label{eq:eomNLO}
\end{align}
${\cal D}_\mu$ and ${\cal F}^{\mu\nu}$ are the covariant derivative
and field strength constructed with the classical field ${\cal
  A}^\mu$. Eq.~(\ref{eq:prop-back}) can also be written as a Gaussian
average over random fields,
\begin{equation}
{\cal G}^{\mu a, \nu b}(x,y)\empile{=}\over{(x-y)^2\le
  0}
\left<A^{\mu a}(x) A^{\nu b}(y)\right>
\label{eq:prop-back-1}
\end{equation}
with
\begin{eqnarray}
&&
A^{\mu a}(x)\equiv \sum_{\lambda,c}\int\frac{d^3\k}{(2\pi)^3 2|\k|} \; \Big[
c_{\k\lambda c}\,a^{\mu a}_{\k\lambda c}(x)+\mbox{c.c.}\Big]\; ,\nonumber\\
&&\big<c_{\k\lambda c}\big>=0\;,\quad \big<c_{\k\lambda c}c_{\k'\lambda' c'}^*\big>
=\frac{1}{2}\;2\big|\k\big|\,(2\pi)^3\delta(\k-\k')\,\delta_{\lambda\lambda'}\,\delta_{cc'}\; .
\label{eq:RF}
\end{eqnarray}
The fields $A^{\mu a}$ in Eq.~(\ref{eq:prop-back-1}) are linear
superpositions of the $a_{\k\lambda c}$ with random coefficients that
are a classical analogue of creation and annihilation operators. Their
variance in Eq.~(\ref{eq:RF}) indicates that they correspond to an
occupation number $1/2$, that can be interpreted as the zero-point
{quantum fluctuations} in each mode.

Using Eq.~(\ref{eq:prop-back}), the NLO correction to an inclusive
observable can be written as follows:
\begin{equation}
\big<{\cal O}\big>_{_{\rm NLO}}
=
\Bigg[
 \frac{1}{2}\!\!\!\!
\int\limits_{_{\u,\v\in\Sigma}}
\!\!\!\!
\int_\k
\big[{\colorb a_\k}\,{\mathbbm T}\big]_\u 
\big[{\colorb a_\k^*}\,{\mathbbm T}\big]_\v
+\int\limits_{_{\u\in\Sigma}}
\big[{\colorb{\bs\alpha}}\,{\mathbbm T}\big]_\u
\Bigg]\;
\big<{\cal O}\big>_{_{\rm LO}}\; .
\label{eq:O-NLO}
\end{equation}
In this formula, $\big<{\cal O}\big>_{_{\rm LO}}$ depends on the
classical field ${\cal A}^\mu$ on a surface $\Sigma$, where the
initial value of the classical field is specified, e.g. a surface
$\tau={\rm const}$. ${\mathbbm T}_\u$ is the generator of shifts of
this initial field: for any function ${\cal F}[{\cal A}_{_\Sigma}]$ of
the fields on $\Sigma$, we have
\begin{equation}
\left[\exp \int_{\u\in\Sigma}[{\bs \alpha}{\mathbbm T}]_\u\right]
\;{\cal F}[{\cal A}_{_\Sigma}]
=
{\cal F}[{\cal A}_{_\Sigma}+{\bs\alpha}]\; .
\end{equation}
The functions $a_\k$ in Eq.~(\ref{eq:O-NLO}) are the mode functions
introduced in Eq.~(\ref{eq:prop-back}) and the function ${\bs \alpha}$
can also be expressed in terms of the same mode
functions. Eq.~(\ref{eq:O-NLO}) shows that the NLO can be obtained
from the LO by fiddling with the initial values of the classical
fields, while the fields continue to evolve classically.  Note that
formulas such as (\ref{eq:O-NLO}) cannot be exact at 2-loops
\cite{Gelis:2012ct}.

\subsection{Boost invariant fluctuations and factorization}
Some of these fluctuations leads to logarithms of the cutoff that
separates the sources from the field degrees of freedom
\cite{Gelis:2008rw,Gelis:2008ad,Gelis:2008sz}. One can
prove that
\begin{eqnarray}
&&
\frac{1}{2}\!\!\!\!
\int\limits_{_{\u,\v\in\Sigma}}
\!\!\!\!
\int_\k
\big[{\colorb a_\k}\,{\mathbbm T}\big]_\u 
\big[{\colorb a_\k^*}\,{\mathbbm T}\big]_\v
+\int\limits_{_{\u\in\Sigma}}
\big[{\colorb{\bs\alpha}}\,{\mathbbm T}\big]_\u
=\nonumber\\
&&\qquad\qquad=
\log(\Lambda^+)\;{\cal H}_1 +\log(\Lambda^-)\;{\cal H}_2\; +\; \mbox{terms w/o logs}\; ,
\label{eq:logs}
\end{eqnarray}
where the operators ${\cal H}_{1,2}$ are the \emph{JIMWLK
  Hamiltonians} of the projectiles \cite{Balitsky:1995ub}.  The
fluctuations that give logarithms of $\Lambda^+$ (resp. $\Lambda^-$)
have a large momentum rapidity in the direction of the nucleus $1$
(resp. nucleus $2$). To the observer, these field fluctuations appear
as fast moving color charges, similar the degrees of freedom already
included in $\rho_{1,2}$.

Eq.~(\ref{eq:logs}) substantiates this intuitive picture. It implies
that these logarithms can be absorbed into redefinitions of the
distributions $W[\rho_{1,2}]$, evolving with the  cutoff according to
(\cite{Balitsky:1995ub,JalilianMarian:1996xn,JalilianMarian:1997jx,JalilianMarian:1997gr,JalilianMarian:1997dw,JalilianMarian:1998cb,Iancu:2000hn,Iancu:2001ad,Ferreiro:2001qy},
see also
\cite{Balitsky:2008zza,Balitsky:2009xg,Balitsky:2013fea,Grabovsky:2013mba,Kovner:2013ona,Kovner:2014xia,Kovner:2014lca}
for next-to-leading log corrections)
\begin{equation}
\frac{\partial W}{\partial \Lambda} = -{\cal H}\; W\; .
\label{eq:JIMWLK}
\end{equation}
This works because Eq.~(\ref{eq:logs}) does not mix $\rho_1$ and
$\rho_2$, a consequence of \emph{causality}: the logarithms come from
soft radiation by fast color charges, which has a long formation time
and must occur long before the collision. Therefore, the radiation in
nuclei 1 and 2 are independent. All the fluctuation modes that have a
rapidity separation $\Delta y\gtrsim \alpha_s^{-1}$ with the observer
can be resummed in this way, leading to a factorized expression
\begin{equation}
\big<{\cal O}\big>\empile{=}\over{\mbox{\scriptsize Leading Log}}
\int \big[D\rho_1\big]\big[D\rho_2\big]\;W_1[\rho_1]\,W_2[\rho_2]\;
{\cal O}_{_{\rm LO}}(\rho_{1,2})\; ,
\end{equation}
where $W_1$ and $W_2$ are solutions of Eq.~(\ref{eq:JIMWLK}), starting
from an initial condition at the rapidity of the projectile and
evolving to the rapidity of the observer. In this new light, the
Gaussian distribution introduced in Eq.~(\ref{eq:MVdist}) should be
viewed as a model for the initial color distribution, but non-Gaussian
correlations may develop when evolving away from the projectile.

Thanks to its structure, the JIMWLK Hamiltonian can be interpreted as
a diffusion operator in a functional space, which allows to reproduce
the rapidity evolution of $W[\rho]$ by a random walk described by a
Langevin equation \cite{Blaizot:2002np}. This approach has been used
in several works addressing the numerical study of the JIMWLK
equation \cite{Rummukainen:2003ns,Dumitru:2011vk}. Note also that, for
simple correlators of color charges, this evolution is well
described by a mean field approximation known as the
\emph{Balitsky-Kovchegov equation} \cite{Kovchegov:1999yj} (see \cite{Balitsky:2006wa,Balitsky:2008zza,Kovchegov:2006vj,Gardi:2006rp} for next-to-leading log corrections).

\subsection{Rapidity dependent fluctuations}
The fluctuation modes that give logarithms in Eq.~(\ref{eq:logs})
preserve the $\eta$-independence of the LO. The remaining modes are
$\eta$-dependent, but do not give large logarithms. It is convenient
to introduce a new basis:
\begin{equation}
b_{\k_\perp\nu\lambda c}^{\mu a}\equiv \int dy\; e^{i\nu y}\; a_{\k_\perp k_z \lambda c}^{\mu a}\; ,
\label{eq:nu-def}
\end{equation}
where $y$ is the momentum rapidity
$y=\ln((k_0+k_z)/(k_0-k_z))/2$. Since the $a_{\k\lambda c}(\x)$ depend
on the momentum and spacetime rapidities only via the difference
$y-\eta$, $b_{\k_\perp\nu\lambda c}$ has a trivial rapidity dependence
in $\exp(i\nu\eta)$.

The mode functions $b_{\k\lambda c}$ can be calculated analytically up
to a proper time $Q_s\tau\ll1$ \cite{Epelbaum:2013waa} (beyond this
time, the classical background field itself is not known
analytically), in $A^\tau=0$ gauge:
\begin{eqnarray}
b^{i a}_{\k_\perp \nu\lambda c}(\tau,\eta,\x)
&=&
F_{\k_\perp\nu\lambda c}^{+,ia}(\tau,\eta,\x)+F_{\k_\perp\nu\lambda c}^{-,ia}(\tau,\eta,\x)
\nonumber\\
b_{\k_\perp \nu\lambda c}^{\eta a}(\tau,\eta,\x)
&=&
{\cal D}^{i}_{ab}\; 
\Big(
\frac{F_{\k_\perp\nu\lambda c}^{+,ib}(\tau,\eta,\x)}{2+i\nu}
-
\frac{F_{\k_\perp\nu\lambda c}^{-,ib}(\tau,\eta,\x)}{2-i\nu}
\Big)\; ,
\nonumber\\
&&
\label{eq:finalresult}
\end{eqnarray}
where we denote
\begin{eqnarray}
&&
F_{\k_\perp\nu\lambda c}^{+,ia}(\tau,\eta,\x)
\equiv
\Gamma(-i\nu)\,e^{+\frac{\nu\pi}{2}}
e^{i\nu\eta}\,{V}^{\dagger}_{1ab}(\x)\,\vartheta^j_{{\bm k} \lambda}
\nonumber\\
&&\quad\times
\int \frac{d^2\p_\perp}{(2\pi)^2}\; e^{i\p\cdot\x}\;
{\widetilde{V}}_{1bc}(\p+\k_\perp)
\left(\frac{p_\perp^2\tau}{2k_\perp}\right)^{+i\nu}
\Big[\delta^{ji} -\frac{2p^j_\perp p^i_\perp}{p^2_\perp}\Big]
\nonumber\\
&&
F_{\k_\perp\nu\lambda c}^{-,ia}(\tau,\eta,\x)
\equiv
\Gamma(+i\nu)\,e^{-\frac{\nu\pi}{2}}
e^{i\nu\eta}\,{V}^{\dagger}_{2ab}(\x)\,
\vartheta^j_{{\bm k} \lambda}
\nonumber\\
&&\quad\times
\int \frac{d^2\p_\perp}{(2\pi)^2}\; e^{i\p\cdot\x}\;
{\widetilde{V}}_{2bc}(\p+\k_\perp)
\left(\frac{p_\perp^2\tau}{2k_\perp}\right)^{-i\nu}
\Big[\delta^{ji} -\frac{2p^j_\perp p^i_\perp}{p^2_\perp}\Big]
\label{eq:F2}
\end{eqnarray}
and
$\vartheta^i_{\k\lambda}\equiv(\delta^{ij}-2\tfrac{k^ik^j}{k_\perp^2})\,\epsilon_{\k\lambda}^j$. The
Wilson lines $V_{1,2}$ have been introduced in
Eq.~(\ref{eq:wilson-def}) (here, they are in the adjoint
representation), and the covariant derivative ${\cal D}^{i}_{ab}$ is
constructed from the initial field $A_0^i$ given in
Eq.~(\ref{eq:init-LO-tau0}). The corresponding electrical fields are
obtained by time derivatives of Eqs.~(\ref{eq:finalresult}).
Since these expressions are only valid at very short proper times,
they should be used as initial conditions for Eq.~(\ref{eq:eomNLO}).

\subsection{Instabilities and classical statistical approximation}
After having resummed the large logarithms via the JIMWLK evolution,
the remaining contributions are seemingly suppressed by a factor
$\alpha_s$ compared to LO. However, this conclusion is invalidated by
the chaotic behavior of classical solutions of Yang-Mills
equations\footnote{These instabilities are related to the Weibel
  instability that happens in anisotropic
  plasmas~\cite{Mrowczynski:1993qm,Mrowczynski:1996vh}.}, that are
exponentially sensitive to their initial conditions
\cite{Biro:1993qc,Heinz:1996wx,Bolte:1999th,Romatschke:2005pm,Romatschke:2005ag,Romatschke:2006nk,Fukushima:2007ja,Fujii:2008dd,Fujii:2009kb,Kunihiro:2010tg,Fukushima:2011nq}. Consequently,
some of the mode functions $a_{\k\lambda c}$ grow exponentially with
time, and the NLO corrections contain exponentially growing terms: the
suppression factor $\alpha_s$ quickly becomes irrelevant in view of
this time dependence.

The fastest growing can be summed to all orders by exponentiating the
quadratic part of the operator that appears in Eq.~(\ref{eq:O-NLO}),
giving an expression that depends on classical fields averaged over a
Gaussian ensemble of fluctuating initial conditions
\cite{Gelis:2007kn}. This approximation is known as \emph{classical
  statistical approximation} (CSA). A strict application of this
scheme leads to initial fields that are the sum of the LO classical
field ${\cal A}^{\mu}$ (non fluctuating), and a fluctuating part given
by Eq.~(\ref{eq:RF}) \cite{Gelis:2013rba}:
\begin{eqnarray}
&&
A^{\mu a}(x)={\cal A}^{\mu a}(x)+ \sum_{\lambda,c}\int\frac{d^3\k}{(2\pi)^3 2|\k|} \; \Big[
c_{\k\lambda c}\,a^{\mu a}_{\k\lambda c}(x)+\mbox{c.c.}\Big]\; ,\nonumber\\
&&\big<c_{\k\lambda c}\big>=0\;,\quad \big<c_{\k\lambda c}c_{\k'\lambda' c'}^*\big>
=\frac{1}{2}\;2\big|\k\big|\,(2\pi)^3\delta(\k-\k')\,\delta_{\lambda\lambda'}\,\delta_{cc'}\; .
\label{eq:RF1}
\end{eqnarray}
However, such a spectrum of fluctuations combined with classical time
evolution leads to results that are very sensitive to the ultraviolet
cutoff \cite{Berges:2013lsa}. More formally, this resummation breaks
the renormalizability of the underlying theory
\cite{Epelbaum:2014yja,Epelbaum:2014mfa}.

A variant of the CSA uses as initial condition the ensemble of fields
corresponding to a classical gas of gluons with a distribution
$f(\k)$. Such initial fields have no non-fluctuating part, and the
variance of the random coefficients is proportional to $f(\k)$
\cite{Berges:2013eia,Berges:2013fga,Berges:2014bba}:
\begin{eqnarray}
&&
A^{\mu a}(x)= \sum_{\lambda,c}\int\frac{d^3\k}{(2\pi)^3 2|\k|} \; \Big[
c_{\k\lambda c}\,a^{\mu a}_{\k\lambda c}(x)+\mbox{c.c.}\Big]\; ,\nonumber\\
&&\big<c_{\k\lambda c}\big>=0\;,\quad \big<c_{\k\lambda c}c_{\k'\lambda' c'}^*\big>
=f(\k)\;2\big|\k\big|\,(2\pi)^3\delta(\k-\k')\,\delta_{\lambda\lambda'}\,\delta_{cc'}\; .
\label{eq:RF2}
\end{eqnarray}
If $f(\k)$ decreases sufficiently fast, this modification leads to
ultraviolet finite results.

\subsection{Quantum fluctuations in kinetic theory}
Eqs.~(\ref{eq:RF1}) and (\ref{eq:RF2}) describe very
different systems, despite their similarity:
\begin{itemize}
\item The flat spectrum of Eq.~(\ref{eq:RF1}) on top of a classical
  field describes a quantum coherent state.
\item The compact spectrum of Eq.~(\ref{eq:RF2}) describes an
  incoherent classical state (quantum mechanics imposes a minimal
  variance $1/2$ in all modes).
\end{itemize}
When applied to simulations of the early stages of heavy ion
collisions, these two types of fluctuating initial conditions lead to
quite different behaviors of the pressure tensor.

In order to clarify the situation, it would be highly desirable to
include quantum fluctuations without sacrificing renormalizability.
In field theory, a framework that achieves this is the
\emph{2-particle irreducible} approximation
\cite{Luttinger:1960ua,Aarts:2002dj,Calzetta:2002ub,Berges:2004pu,Berges:2004yj}
of the Kadanoff-Baym equations \cite{Baym:1961zz}. Although in
principle feasible even for the expanding geometry encountered in
heavy ion collisions \cite{Hatta:2011ky,Hatta:2012gq}, this is much
more complicated to implement than the CSA.

A simpler alternative is to study quantum fluctuations in kinetic
theory. Schematically, the Boltzmann equation for $2\to 2$ scatterings
reads:
\begin{eqnarray}
\partial_t f_3
&\sim&
g^4\int_{124}
\cdots
\big[f_1f_2({f_3+f_4})-{f_3f_4}(f_1+f_2)\big]
\nonumber\\
&&\quad+g^4\int_{124}\cdots\big[f_1f_2-{f_3f_4}\big]\; .
\label{eq:boltz}
\end{eqnarray}
where the dots encapsulates the cross-section and the delta functions
for energy-momentum conservation, whose details are irrelevant here.
In Eq.~(\ref{eq:boltz}), we have written on the first line the terms
that correspond to the classical approximation of Eq.~(\ref{eq:RF2}),
and on the second line the terms that come from quantum fluctuations.

Keeping only the cubic terms of the first line may be justified when
the occupation number is large.  However, this approximation is not
uniform over all momentum space, which may be especially problematic
in systems with an anisotropic momentum distribution.
\begin{figure}[htbp]
\begin{center}
 \begin{minipage}{0.25\textwidth}
   \includegraphics[width=\textwidth]{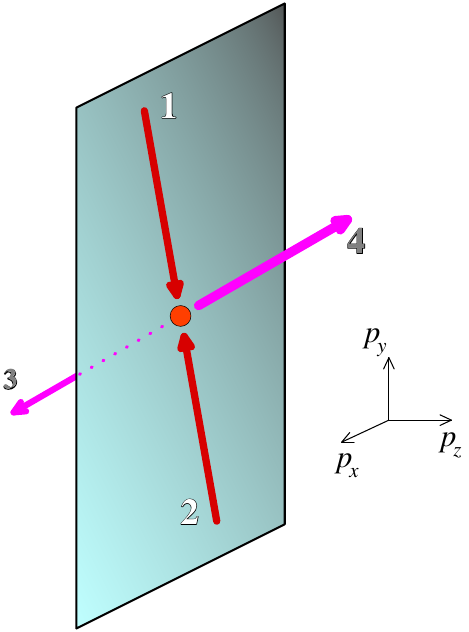}
 \end{minipage}\hspace{1cm}
 \begin{minipage}{0.5\textwidth}
   \includegraphics[width=\textwidth]{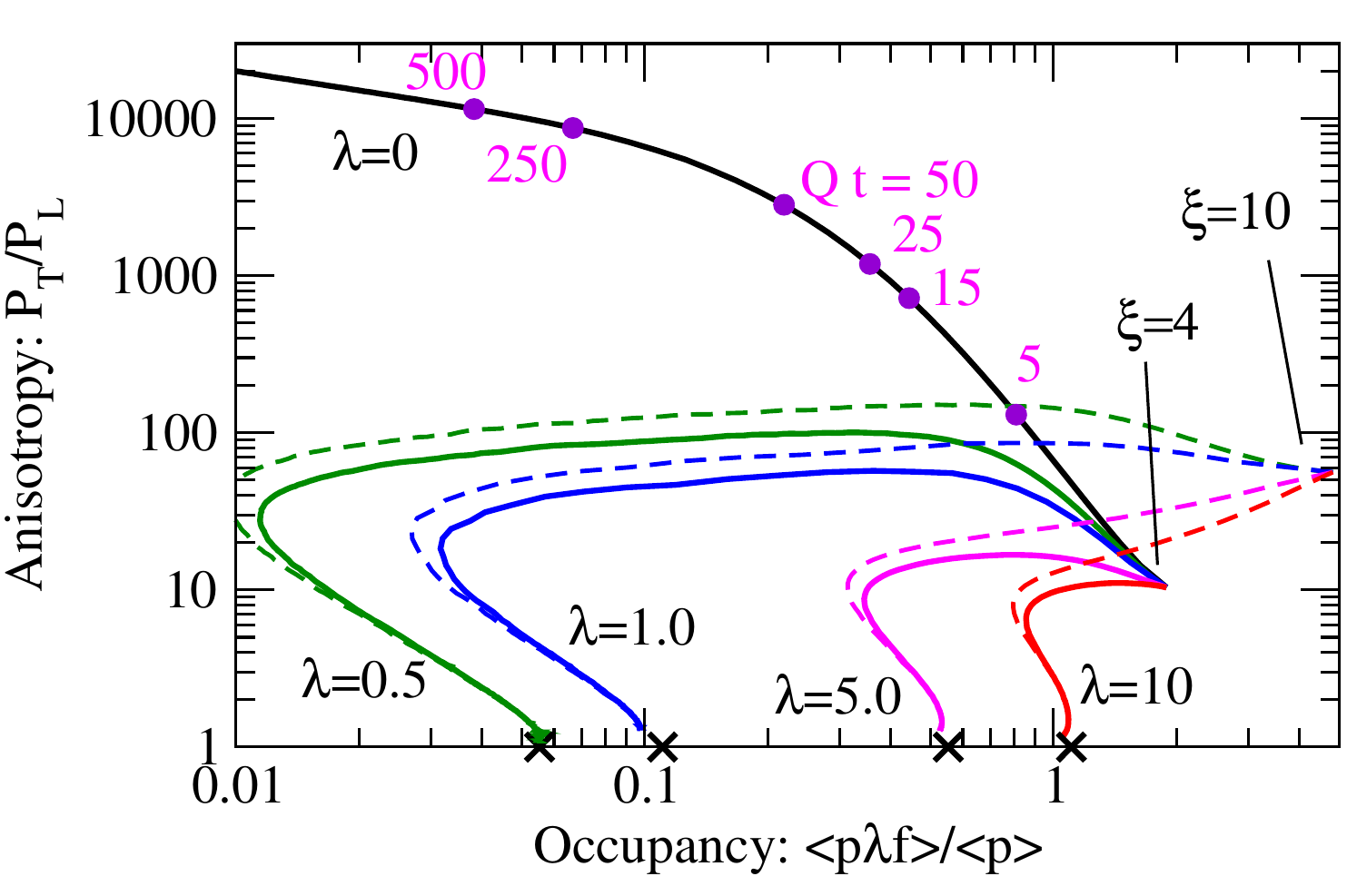}
  \end{minipage}
\end{center}
\caption{\label{fig:iso}Left: $2\to 2$ scattering contributing to
  isotropization. Right: time evolution of the anisotropy with the
  classical ($\lambda=0$) and complete ($\lambda=0.5$ to $10$)
  Boltzmann equations (plot taken from \cite{Kurkela:2015qoa}, with
  time labels added by us for clarity).}
\end{figure}
In the left part of Fig.~\ref{fig:iso}, we illustrate this for a
distribution nearly proportional to $\delta(p_z)$, for which the
incoming particles 1,2 have purely transverse momenta. Isotropization
requires that at least one of the outgoing particles (3 or 4) has a
nonzero $p_z$. Nonzero contributions with this kinematics can only
come from the second line, which is dropped in the classical
approximation.

This has been seen in numerical studies of the Boltzmann equation for
a longitudinally expanding
system~\cite{Epelbaum:2015vxa,Kurkela:2015qoa}. The right part of
Fig.\,\ref{fig:iso} shows results with (curves $\lambda=0.5$ to $10$)
and without (curve $\lambda=0$) the terms of the second line. Starting
at the time $Q\tau=1$ from identical CGC-like initial conditions, the
classical and quantum evolutions starts diverging around $Q\tau\approx
2$ for $\lambda=0.5$ (i.e. $\alpha_s\approx 0.02$ for $N_c=2$ colors),
well before the conjectured range of validity of the classical
approximation, $Q\tau\approx \alpha_s^{-3/2}\approx 350$. These
considerations show that quantum fluctuations are essential for
isotropization: purely classical approximations do not capture the
relevant physics and fail at assessing their own range of validity.

\section{Fluctuations in small systems, proton shape}\label{sec:small}
Proton-nucleus collisions, which where initially thought to be an easy
to understand reference for heavy ion collisions, have lead to a number of new puzzles since a lot of the collectivity that
appears in bigger systems is also visible there. While in
nucleus-nucleus collisions, the fluctuations that are internal to a
nucleon do not play a big role, because they are averaged over many
nucleons, they may become crucial in proton-nucleus collisions.

The fluctuating geometry of a proton will most affect final
observables such as anisotropy coefficients $v_n$ if final state
collective effects are dominant.  
Hydrodynamic calculations using fluctuating Monte Carlo Glauber initial conditions
agree with a variety of experimental observables
\cite{Bozek:2011if,Bozek:2012gr,Bozek:2013ska,Werner:2013tya,Werner:2013ipa,Bozek:2013uha,Schenke:2014zha,Kozlov:2014fqa}.
However, more sophisticated models such as the IP-Glasma
could not describe the measured $v_2$ and $v_3$ coefficients in p+Pb collisions \cite{Schenke:2014zha}. 
The reason for this disagreement was the assumption of a round proton, which was not important 
for heavy ion collisions.
However, in small systems additional sub-nucleonic fluctuations can make a significant difference. 
For example, the initial color charges in a proton could be concentrated around three
valence quarks. It was shown in \cite{Schlichting:2014ipa} that
a proton will retain a memory of the
fluctuating shape at large $x$ after JIMWLK evolution over several
units in rapidity. This means that a proton at high energy can have
fluctuations on various length scales, not just $1/Q_s$.

An issue with the use of hydrodynamics in small systems is the
possible break-down of its applicability. 
For small systems the Knudsen number, the
ratio of a typical microscopic over a macroscopic scale, can become large,
indicating that the hydrodynamic framework is beginning to
break down \cite{Niemi:2014wta}. However, this does not preclude
the existence of final state effects, which could be described within a framework other than hydrodynamics.

Apart from the possibility that final state collective effects are
dominant also in small collision systems, like p+A, initial state
correlations that affect particle production can also contribute to
the measured anisotropies
\cite{Kovner:2010xk,Dusling:2013qoz,Dumitru:2014yza,Lappi:2015vha,Lappi:2015vta}. In
particular, the classical Yang Mills framework, which is the basis for
the IP-Glasma model, contains multi gluon correlations that show
qualitatively similar features as the experimental data, namely a
relatively large $v_2$ and $v_3$ in p+A collisions
\cite{Schenke:2015aqa}. Such effects would naturally lead to the
observed long range rapidity correlations, which have to be assumed in
most hydrodynamic calculations.

At this point it is not yet settled which of the two distinct effects
described above dominates the creation of the observed anisotropies. A
detailed discussion on the current status of the field can be found in
\cite{Dusling:2015gta}. Here we conclude that in either case the
detailed understanding of sub-nucleonic fluctuations in the initial state is essential for the
interpretation of the experimental data of multi particle correlations
in small collision systems.

\section{Conclusions}
Fluctuations play a very important role at various levels in the
description of heavy ion collisions. Phenomenologically, fluctuations
are very important in obtaining the correct final state azimuthal
correlations. Their effect is comparatively larger in collisions
involving smaller observables, and it is still an open question
whether the azimuthal patterns observed in the collision of small
systems (p-p collisions) are mostly due to initial state quantum
fluctuations.

On a more theoretical level, quantum fluctuations also seem essential
in explaining the isotropization of the pressure tensor and the early
applicability of hydrodynamics.  The complete description of the early
time dynamics and transition to the hydrodynamic regime is still an
active field of research. Likely the full answer will involve various
stages including the unstable evolution in classical Yang-Mills
dynamics with quantum corrections and kinetic theory.

\section*{Acknowledgements}
This manuscript has been accepted by Annual Reviews of Nuclear and Particle Science.
FG is supported by the Agence Nationale de la Recherche project
11-BS04-015-01. BPS is supported by the US Department of Energy under
DOE Contract No.  DE-SC0012704.

\bibliography{spires}

\begin{thebibliography}{144}
\expandafter\ifx\csname natexlab\endcsname\relax\def\natexlab#1{#1}\fi
\expandafter\ifx\csname bibnamefont\endcsname\relax
  \def\bibnamefont#1{#1}\fi
\expandafter\ifx\csname bibfnamefont\endcsname\relax
  \def\bibfnamefont#1{#1}\fi
\expandafter\ifx\csname citenamefont\endcsname\relax
  \def\citenamefont#1{#1}\fi
\expandafter\ifx\csname url\endcsname\relax
  \def\url#1{\texttt{#1}}\fi
\expandafter\ifx\csname urlprefix\endcsname\relax\def\urlprefix{URL }\fi
\providecommand{\bibinfo}[2]{#2}
\providecommand{\eprint}[2][]{\url{#2}}

\bibitem[{\citenamefont{Hirano and Tsuda}(2002)}]{Hirano:2002ds}
\bibinfo{author}{\bibfnamefont{T.}~\bibnamefont{Hirano}} \bibnamefont{and}
  \bibinfo{author}{\bibfnamefont{K.}~\bibnamefont{Tsuda}},
  \bibinfo{journal}{Phys. Rev.} \textbf{\bibinfo{volume}{C66}},
  \bibinfo{pages}{054905} (\bibinfo{year}{2002}).

\bibitem[{\citenamefont{Kolb and Heinz}(2003)}]{Kolb:2003dz}
\bibinfo{author}{\bibfnamefont{P.~F.} \bibnamefont{Kolb}} \bibnamefont{and}
  \bibinfo{author}{\bibfnamefont{U.~W.} \bibnamefont{Heinz}},
  \bibinfo{journal}{Quark Gluon Plasma 3 ed R C Hwa and X N Wang (Singapore:
  World Scientific)} p. \bibinfo{pages}{634} (\bibinfo{year}{2003}),
  \eprint{nucl-th/0305084}.

\bibitem[{\citenamefont{Huovinen}(2003)}]{Huovinen:2003fa}
\bibinfo{author}{\bibfnamefont{P.}~\bibnamefont{Huovinen}},
  \bibinfo{journal}{Quark Gluon Plasma 3 ed R C Hwa and X N Wang (Singapore:
  World Scientific)} p. \bibinfo{pages}{600} (\bibinfo{year}{2003}),
  \eprint{nucl-th/0305064}.

\bibitem[{\citenamefont{Romatschke and Romatschke}(2007)}]{Romatschke:2007mq}
\bibinfo{author}{\bibfnamefont{P.}~\bibnamefont{Romatschke}} \bibnamefont{and}
  \bibinfo{author}{\bibfnamefont{U.}~\bibnamefont{Romatschke}},
  \bibinfo{journal}{Phys. Rev. Lett.} \textbf{\bibinfo{volume}{99}},
  \bibinfo{pages}{172301} (\bibinfo{year}{2007}).

\bibitem[{\citenamefont{Policastro et~al.}(2001)\citenamefont{Policastro, Son,
  and Starinets}}]{Policastro:2001yc}
\bibinfo{author}{\bibfnamefont{G.}~\bibnamefont{Policastro}},
  \bibinfo{author}{\bibfnamefont{D.~T.} \bibnamefont{Son}}, \bibnamefont{and}
  \bibinfo{author}{\bibfnamefont{A.~O.} \bibnamefont{Starinets}},
  \bibinfo{journal}{Phys. Rev. Lett.} \textbf{\bibinfo{volume}{87}},
  \bibinfo{pages}{081601} (\bibinfo{year}{2001}), \eprint{hep-th/0104066}.

\bibitem[{\citenamefont{Kovtun et~al.}(2005)}]{Kovtun:2004de}
\bibinfo{author}{\bibfnamefont{P.}~\bibnamefont{Kovtun}} \bibnamefont{et~al.},
  \bibinfo{journal}{Phys. Rev. Lett.} \textbf{\bibinfo{volume}{94}},
  \bibinfo{pages}{111601} (\bibinfo{year}{2005}).

\bibitem[{\citenamefont{Schenke et~al.}(2011)\citenamefont{Schenke, Jeon, and
  Gale}}]{Schenke:2010rr}
\bibinfo{author}{\bibfnamefont{B.}~\bibnamefont{Schenke}},
  \bibinfo{author}{\bibfnamefont{S.}~\bibnamefont{Jeon}}, \bibnamefont{and}
  \bibinfo{author}{\bibfnamefont{C.}~\bibnamefont{Gale}},
  \bibinfo{journal}{Phys. Rev. Lett.} \textbf{\bibinfo{volume}{106}},
  \bibinfo{pages}{042301} (\bibinfo{year}{2011}).

\bibitem[{\citenamefont{Bozek and Broniowski}(2012)}]{Bozek:2012fw}
\bibinfo{author}{\bibfnamefont{P.}~\bibnamefont{Bozek}} \bibnamefont{and}
  \bibinfo{author}{\bibfnamefont{W.}~\bibnamefont{Broniowski}},
  \bibinfo{journal}{Phys. Rev.} \textbf{\bibinfo{volume}{C85}},
  \bibinfo{pages}{044910} (\bibinfo{year}{2012}), \eprint{1203.1810}.

\bibitem[{\citenamefont{Karpenko et~al.}(2015)\citenamefont{Karpenko, Huovinen,
  Petersen, and Bleicher}}]{Karpenko:2015xea}
\bibinfo{author}{\bibfnamefont{I.~A.} \bibnamefont{Karpenko}},
  \bibinfo{author}{\bibfnamefont{P.}~\bibnamefont{Huovinen}},
  \bibinfo{author}{\bibfnamefont{H.}~\bibnamefont{Petersen}}, \bibnamefont{and}
  \bibinfo{author}{\bibfnamefont{M.}~\bibnamefont{Bleicher}},
  \bibinfo{journal}{Phys. Rev.} \textbf{\bibinfo{volume}{C91}},
  \bibinfo{pages}{064901} (\bibinfo{year}{2015}).

\bibitem[{\citenamefont{Heinz and Snellings}(2013)}]{Heinz:2013th}
\bibinfo{author}{\bibfnamefont{U.}~\bibnamefont{Heinz}} \bibnamefont{and}
  \bibinfo{author}{\bibfnamefont{R.}~\bibnamefont{Snellings}},
  \bibinfo{journal}{Ann. Rev. Nucl. Part. Sci.} \textbf{\bibinfo{volume}{63}},
  \bibinfo{pages}{123} (\bibinfo{year}{2013}), \eprint{1301.2826}.

\bibitem[{\citenamefont{Gale et~al.}(2013{\natexlab{a}})\citenamefont{Gale,
  Jeon, and Schenke}}]{Gale:2013da}
\bibinfo{author}{\bibfnamefont{C.}~\bibnamefont{Gale}},
  \bibinfo{author}{\bibfnamefont{S.}~\bibnamefont{Jeon}}, \bibnamefont{and}
  \bibinfo{author}{\bibfnamefont{B.}~\bibnamefont{Schenke}},
  \bibinfo{journal}{Int. J. Mod. Phys.} \textbf{\bibinfo{volume}{A28}},
  \bibinfo{pages}{1340011} (\bibinfo{year}{2013}{\natexlab{a}}),
  \eprint{1301.5893}.

\bibitem[{\citenamefont{de~Souza et~al.}(2015)\citenamefont{de~Souza, Koide,
  and Kodama}}]{deSouza:2015ena}
\bibinfo{author}{\bibfnamefont{R.~D.} \bibnamefont{de~Souza}},
  \bibinfo{author}{\bibfnamefont{T.}~\bibnamefont{Koide}}, \bibnamefont{and}
  \bibinfo{author}{\bibfnamefont{T.}~\bibnamefont{Kodama}}
  (\bibinfo{year}{2015}), \eprint{1506.03863}.

\bibitem[{\citenamefont{Jeon and Heinz}(2015)}]{Jeon:2015dfa}
\bibinfo{author}{\bibfnamefont{S.}~\bibnamefont{Jeon}} \bibnamefont{and}
  \bibinfo{author}{\bibfnamefont{U.}~\bibnamefont{Heinz}},
  \bibinfo{journal}{Int. J. Mod. Phys.} \textbf{\bibinfo{volume}{E24}},
  \bibinfo{pages}{1530010} (\bibinfo{year}{2015}), \eprint{1503.03931}.

\bibitem[{\citenamefont{Alver et~al.}(2007)}]{Alver:2006wh}
\bibinfo{author}{\bibfnamefont{B.}~\bibnamefont{Alver}} \bibnamefont{et~al.}
  (\bibinfo{collaboration}{PHOBOS}), \bibinfo{journal}{Phys. Rev. Lett.}
  \textbf{\bibinfo{volume}{98}}, \bibinfo{pages}{242302}
  (\bibinfo{year}{2007}), \eprint{nucl-ex/0610037}.

\bibitem[{\citenamefont{Mishra et~al.}(2008)\citenamefont{Mishra, Mohapatra,
  Saumia, and Srivastava}}]{Mishra:2007tw}
\bibinfo{author}{\bibfnamefont{A.~P.} \bibnamefont{Mishra}},
  \bibinfo{author}{\bibfnamefont{R.~K.} \bibnamefont{Mohapatra}},
  \bibinfo{author}{\bibfnamefont{P.~S.} \bibnamefont{Saumia}},
  \bibnamefont{and} \bibinfo{author}{\bibfnamefont{A.~M.}
  \bibnamefont{Srivastava}}, \bibinfo{journal}{Phys. Rev.}
  \textbf{\bibinfo{volume}{C77}}, \bibinfo{pages}{064902}
  (\bibinfo{year}{2008}), \eprint{0711.1323}.

\bibitem[{\citenamefont{Alver and Roland}(2010)}]{Alver:2010gr}
\bibinfo{author}{\bibfnamefont{B.}~\bibnamefont{Alver}} \bibnamefont{and}
  \bibinfo{author}{\bibfnamefont{G.}~\bibnamefont{Roland}},
  \bibinfo{journal}{Phys. Rev.} \textbf{\bibinfo{volume}{C81}},
  \bibinfo{pages}{054905} (\bibinfo{year}{2010}).

\bibitem[{\citenamefont{Alver et~al.}(2010)\citenamefont{Alver, Gombeaud,
  Luzum, and Ollitrault}}]{Alver:2010dn}
\bibinfo{author}{\bibfnamefont{B.~H.} \bibnamefont{Alver}},
  \bibinfo{author}{\bibfnamefont{C.}~\bibnamefont{Gombeaud}},
  \bibinfo{author}{\bibfnamefont{M.}~\bibnamefont{Luzum}}, \bibnamefont{and}
  \bibinfo{author}{\bibfnamefont{J.-Y.} \bibnamefont{Ollitrault}},
  \bibinfo{journal}{Phys.Rev.} \textbf{\bibinfo{volume}{C82}},
  \bibinfo{pages}{034913} (\bibinfo{year}{2010}).

\bibitem[{\citenamefont{Aad et~al.}(2013)}]{Aad:2013xma}
\bibinfo{author}{\bibfnamefont{G.}~\bibnamefont{Aad}} \bibnamefont{et~al.}
  (\bibinfo{collaboration}{ATLAS}), \bibinfo{journal}{JHEP}
  \textbf{\bibinfo{volume}{11}}, \bibinfo{pages}{183} (\bibinfo{year}{2013}),
  \eprint{1305.2942}.

\bibitem[{\citenamefont{Niemi et~al.}(2013)\citenamefont{Niemi, Denicol,
  Holopainen, and Huovinen}}]{Niemi:2012aj}
\bibinfo{author}{\bibfnamefont{H.}~\bibnamefont{Niemi}},
  \bibinfo{author}{\bibfnamefont{G.~S.} \bibnamefont{Denicol}},
  \bibinfo{author}{\bibfnamefont{H.}~\bibnamefont{Holopainen}},
  \bibnamefont{and} \bibinfo{author}{\bibfnamefont{P.}~\bibnamefont{Huovinen}},
  \bibinfo{journal}{Phys. Rev.} \textbf{\bibinfo{volume}{C87}},
  \bibinfo{pages}{054901} (\bibinfo{year}{2013}), \eprint{1212.1008}.

\bibitem[{\citenamefont{Schenke
  et~al.}(2012{\natexlab{a}})\citenamefont{Schenke, Tribedy, and
  Venugopalan}}]{Schenke:2012wb}
\bibinfo{author}{\bibfnamefont{B.}~\bibnamefont{Schenke}},
  \bibinfo{author}{\bibfnamefont{P.}~\bibnamefont{Tribedy}}, \bibnamefont{and}
  \bibinfo{author}{\bibfnamefont{R.}~\bibnamefont{Venugopalan}},
  \bibinfo{journal}{Phys. Rev. Lett.} \textbf{\bibinfo{volume}{108}},
  \bibinfo{pages}{252301} (\bibinfo{year}{2012}{\natexlab{a}}).

\bibitem[{\citenamefont{Schenke
  et~al.}(2012{\natexlab{b}})\citenamefont{Schenke, Tribedy, and
  Venugopalan}}]{Schenke:2012hg}
\bibinfo{author}{\bibfnamefont{B.}~\bibnamefont{Schenke}},
  \bibinfo{author}{\bibfnamefont{P.}~\bibnamefont{Tribedy}}, \bibnamefont{and}
  \bibinfo{author}{\bibfnamefont{R.}~\bibnamefont{Venugopalan}},
  \bibinfo{journal}{Phys. Rev.} \textbf{\bibinfo{volume}{C86}},
  \bibinfo{pages}{034908} (\bibinfo{year}{2012}{\natexlab{b}}).

\bibitem[{\citenamefont{Gale et~al.}(2013{\natexlab{b}})\citenamefont{Gale,
  Jeon, Schenke, Tribedy, and Venugopalan}}]{Gale:2012rq}
\bibinfo{author}{\bibfnamefont{C.}~\bibnamefont{Gale}},
  \bibinfo{author}{\bibfnamefont{S.}~\bibnamefont{Jeon}},
  \bibinfo{author}{\bibfnamefont{B.}~\bibnamefont{Schenke}},
  \bibinfo{author}{\bibfnamefont{P.}~\bibnamefont{Tribedy}}, \bibnamefont{and}
  \bibinfo{author}{\bibfnamefont{R.}~\bibnamefont{Venugopalan}},
  \bibinfo{journal}{Phys.Rev.Lett.} \textbf{\bibinfo{volume}{110}},
  \bibinfo{pages}{012302} (\bibinfo{year}{2013}{\natexlab{b}}).

\bibitem[{\citenamefont{Niemi et~al.}(2015)\citenamefont{Niemi, Eskola, and
  Paatelainen}}]{Niemi:2015qia}
\bibinfo{author}{\bibfnamefont{H.}~\bibnamefont{Niemi}},
  \bibinfo{author}{\bibfnamefont{K.~J.} \bibnamefont{Eskola}},
  \bibnamefont{and}
  \bibinfo{author}{\bibfnamefont{R.}~\bibnamefont{Paatelainen}}
  (\bibinfo{year}{2015}), \eprint{1505.02677}.

\bibitem[{\citenamefont{Moreland et~al.}(2015)\citenamefont{Moreland, Bernhard,
  and Bass}}]{Moreland:2014oya}
\bibinfo{author}{\bibfnamefont{J.~S.} \bibnamefont{Moreland}},
  \bibinfo{author}{\bibfnamefont{J.~E.} \bibnamefont{Bernhard}},
  \bibnamefont{and} \bibinfo{author}{\bibfnamefont{S.~A.} \bibnamefont{Bass}},
  \bibinfo{journal}{Phys. Rev.} \textbf{\bibinfo{volume}{C92}},
  \bibinfo{pages}{011901} (\bibinfo{year}{2015}), \eprint{1412.4708}.

\bibitem[{\citenamefont{Martinez and Strickland}(2011)}]{Martinez:2010sd}
\bibinfo{author}{\bibfnamefont{M.}~\bibnamefont{Martinez}} \bibnamefont{and}
  \bibinfo{author}{\bibfnamefont{M.}~\bibnamefont{Strickland}},
  \bibinfo{journal}{Nucl. Phys.} \textbf{\bibinfo{volume}{A856}},
  \bibinfo{pages}{68} (\bibinfo{year}{2011}), \eprint{1011.3056}.

\bibitem[{\citenamefont{Martinez et~al.}(2012)\citenamefont{Martinez,
  Ryblewski, and Strickland}}]{Martinez:2012tu}
\bibinfo{author}{\bibfnamefont{M.}~\bibnamefont{Martinez}},
  \bibinfo{author}{\bibfnamefont{R.}~\bibnamefont{Ryblewski}},
  \bibnamefont{and}
  \bibinfo{author}{\bibfnamefont{M.}~\bibnamefont{Strickland}},
  \bibinfo{journal}{Phys. Rev.} \textbf{\bibinfo{volume}{C85}},
  \bibinfo{pages}{064913} (\bibinfo{year}{2012}), \eprint{1204.1473}.

\bibitem[{\citenamefont{Florkowski et~al.}(2013)\citenamefont{Florkowski,
  Ryblewski, and Strickland}}]{Florkowski:2013lza}
\bibinfo{author}{\bibfnamefont{W.}~\bibnamefont{Florkowski}},
  \bibinfo{author}{\bibfnamefont{R.}~\bibnamefont{Ryblewski}},
  \bibnamefont{and}
  \bibinfo{author}{\bibfnamefont{M.}~\bibnamefont{Strickland}},
  \bibinfo{journal}{Nucl. Phys.} \textbf{\bibinfo{volume}{A916}},
  \bibinfo{pages}{249} (\bibinfo{year}{2013}), \eprint{1304.0665}.

\bibitem[{\citenamefont{Bazow et~al.}(2014)\citenamefont{Bazow, Heinz, and
  Strickland}}]{Bazow:2013ifa}
\bibinfo{author}{\bibfnamefont{D.}~\bibnamefont{Bazow}},
  \bibinfo{author}{\bibfnamefont{U.~W.} \bibnamefont{Heinz}}, \bibnamefont{and}
  \bibinfo{author}{\bibfnamefont{M.}~\bibnamefont{Strickland}},
  \bibinfo{journal}{Phys. Rev.} \textbf{\bibinfo{volume}{C90}},
  \bibinfo{pages}{054910} (\bibinfo{year}{2014}), \eprint{1311.6720}.

\bibitem[{\citenamefont{Strickland}(2014)}]{Strickland:2014pga}
\bibinfo{author}{\bibfnamefont{M.}~\bibnamefont{Strickland}},
  \bibinfo{journal}{Acta Phys. Polon.} \textbf{\bibinfo{volume}{B45}},
  \bibinfo{pages}{2355} (\bibinfo{year}{2014}), \eprint{1410.5786}.

\bibitem[{\citenamefont{Tinti et~al.}(2016)\citenamefont{Tinti, Ryblewski,
  Florkowski, and Strickland}}]{Tinti:2015xra}
\bibinfo{author}{\bibfnamefont{L.}~\bibnamefont{Tinti}},
  \bibinfo{author}{\bibfnamefont{R.}~\bibnamefont{Ryblewski}},
  \bibinfo{author}{\bibfnamefont{W.}~\bibnamefont{Florkowski}},
  \bibnamefont{and}
  \bibinfo{author}{\bibfnamefont{M.}~\bibnamefont{Strickland}},
  \bibinfo{journal}{Nucl. Phys.} \textbf{\bibinfo{volume}{A946}},
  \bibinfo{pages}{29} (\bibinfo{year}{2016}), \eprint{1505.06456}.

\bibitem[{\citenamefont{Nopoush et~al.}(2015)\citenamefont{Nopoush, Strickland,
  Ryblewski, Bazow, Heinz, and Martinez}}]{Nopoush:2015yga}
\bibinfo{author}{\bibfnamefont{M.}~\bibnamefont{Nopoush}},
  \bibinfo{author}{\bibfnamefont{M.}~\bibnamefont{Strickland}},
  \bibinfo{author}{\bibfnamefont{R.}~\bibnamefont{Ryblewski}},
  \bibinfo{author}{\bibfnamefont{D.}~\bibnamefont{Bazow}},
  \bibinfo{author}{\bibfnamefont{U.}~\bibnamefont{Heinz}}, \bibnamefont{and}
  \bibinfo{author}{\bibfnamefont{M.}~\bibnamefont{Martinez}},
  \bibinfo{journal}{Phys. Rev.} \textbf{\bibinfo{volume}{C92}},
  \bibinfo{pages}{044912} (\bibinfo{year}{2015}), \eprint{1506.05278}.

\bibitem[{\citenamefont{Heller et~al.}(2012)\citenamefont{Heller, Janik, and
  Witaszczyk}}]{Heller:2011ju}
\bibinfo{author}{\bibfnamefont{M.~P.} \bibnamefont{Heller}},
  \bibinfo{author}{\bibfnamefont{R.~A.} \bibnamefont{Janik}}, \bibnamefont{and}
  \bibinfo{author}{\bibfnamefont{P.}~\bibnamefont{Witaszczyk}},
  \bibinfo{journal}{Phys. Rev. Lett.} \textbf{\bibinfo{volume}{108}},
  \bibinfo{pages}{201602} (\bibinfo{year}{2012}), \eprint{1103.3452}.

\bibitem[{\citenamefont{Kurkela and Zhu}(2015)}]{Kurkela:2015qoa}
\bibinfo{author}{\bibfnamefont{A.}~\bibnamefont{Kurkela}} \bibnamefont{and}
  \bibinfo{author}{\bibfnamefont{Y.}~\bibnamefont{Zhu}},
  \bibinfo{journal}{Phys. Rev. Lett.} \textbf{\bibinfo{volume}{115}},
  \bibinfo{pages}{182301} (\bibinfo{year}{2015}), \eprint{1506.06647}.

\bibitem[{\citenamefont{Deshpande et~al.}(2009)\citenamefont{Deshpande, Ent,
  and Milner}}]{Deshpande:2009zz}
\bibinfo{author}{\bibfnamefont{A.}~\bibnamefont{Deshpande}},
  \bibinfo{author}{\bibfnamefont{R.}~\bibnamefont{Ent}}, \bibnamefont{and}
  \bibinfo{author}{\bibfnamefont{R.}~\bibnamefont{Milner}},
  \bibinfo{journal}{CERN Cour.} \textbf{\bibinfo{volume}{49N9}},
  \bibinfo{pages}{13} (\bibinfo{year}{2009}).

\bibitem[{\citenamefont{Gelis et~al.}(2010)\citenamefont{Gelis, Iancu,
  Jalilian-Marian, and Venugopalan}}]{Gelis:2010nm}
\bibinfo{author}{\bibfnamefont{F.}~\bibnamefont{Gelis}},
  \bibinfo{author}{\bibfnamefont{E.}~\bibnamefont{Iancu}},
  \bibinfo{author}{\bibfnamefont{J.}~\bibnamefont{Jalilian-Marian}},
  \bibnamefont{and}
  \bibinfo{author}{\bibfnamefont{R.}~\bibnamefont{Venugopalan}},
  \bibinfo{journal}{Ann. Rev. Nucl. Part. Sci.} \textbf{\bibinfo{volume}{60}},
  \bibinfo{pages}{463} (\bibinfo{year}{2010}), \eprint{1002.0333}.

\bibitem[{\citenamefont{McLerran and
  Venugopalan}(1994{\natexlab{a}})}]{McLerran:1994ni}
\bibinfo{author}{\bibfnamefont{L.~D.} \bibnamefont{McLerran}} \bibnamefont{and}
  \bibinfo{author}{\bibfnamefont{R.}~\bibnamefont{Venugopalan}},
  \bibinfo{journal}{Phys. Rev.} \textbf{\bibinfo{volume}{D49}},
  \bibinfo{pages}{2233} (\bibinfo{year}{1994}{\natexlab{a}}).

\bibitem[{\citenamefont{McLerran and
  Venugopalan}(1994{\natexlab{b}})}]{McLerran:1994ka}
\bibinfo{author}{\bibfnamefont{L.~D.} \bibnamefont{McLerran}} \bibnamefont{and}
  \bibinfo{author}{\bibfnamefont{R.}~\bibnamefont{Venugopalan}},
  \bibinfo{journal}{Phys. Rev.} \textbf{\bibinfo{volume}{D49}},
  \bibinfo{pages}{3352} (\bibinfo{year}{1994}{\natexlab{b}}).

\bibitem[{\citenamefont{Kovchegov}(1996)}]{Kovchegov:1996ty}
\bibinfo{author}{\bibfnamefont{Y.~V.} \bibnamefont{Kovchegov}},
  \bibinfo{journal}{Phys. Rev.} \textbf{\bibinfo{volume}{D54}},
  \bibinfo{pages}{5463} (\bibinfo{year}{1996}), \eprint{hep-ph/9605446}.

\bibitem[{\citenamefont{Kovner et~al.}(1995)\citenamefont{Kovner, McLerran, and
  Weigert}}]{Kovner:1995ja}
\bibinfo{author}{\bibfnamefont{A.}~\bibnamefont{Kovner}},
  \bibinfo{author}{\bibfnamefont{L.~D.} \bibnamefont{McLerran}},
  \bibnamefont{and} \bibinfo{author}{\bibfnamefont{H.}~\bibnamefont{Weigert}},
  \bibinfo{journal}{Phys. Rev.} \textbf{\bibinfo{volume}{D52}},
  \bibinfo{pages}{6231} (\bibinfo{year}{1995}), \eprint{hep-ph/9502289}.

\bibitem[{\citenamefont{Krasnitz and Venugopalan}(1999)}]{Krasnitz:1998ns}
\bibinfo{author}{\bibfnamefont{A.}~\bibnamefont{Krasnitz}} \bibnamefont{and}
  \bibinfo{author}{\bibfnamefont{R.}~\bibnamefont{Venugopalan}},
  \bibinfo{journal}{Nucl. Phys.} \textbf{\bibinfo{volume}{B557}},
  \bibinfo{pages}{237} (\bibinfo{year}{1999}), \eprint{hep-ph/9809433}.

\bibitem[{\citenamefont{Krasnitz and Venugopalan}(2000)}]{Krasnitz:1999wc}
\bibinfo{author}{\bibfnamefont{A.}~\bibnamefont{Krasnitz}} \bibnamefont{and}
  \bibinfo{author}{\bibfnamefont{R.}~\bibnamefont{Venugopalan}},
  \bibinfo{journal}{Phys. Rev. Lett.} \textbf{\bibinfo{volume}{84}},
  \bibinfo{pages}{4309} (\bibinfo{year}{2000}).

\bibitem[{\citenamefont{Krasnitz and Venugopalan}(2001)}]{Krasnitz:2000gz}
\bibinfo{author}{\bibfnamefont{A.}~\bibnamefont{Krasnitz}} \bibnamefont{and}
  \bibinfo{author}{\bibfnamefont{R.}~\bibnamefont{Venugopalan}},
  \bibinfo{journal}{Phys. Rev. Lett.} \textbf{\bibinfo{volume}{86}},
  \bibinfo{pages}{1717} (\bibinfo{year}{2001}).

\bibitem[{\citenamefont{Krasnitz et~al.}(2001)\citenamefont{Krasnitz, Nara, and
  Venugopalan}}]{Krasnitz:2001qu}
\bibinfo{author}{\bibfnamefont{A.}~\bibnamefont{Krasnitz}},
  \bibinfo{author}{\bibfnamefont{Y.}~\bibnamefont{Nara}}, \bibnamefont{and}
  \bibinfo{author}{\bibfnamefont{R.}~\bibnamefont{Venugopalan}},
  \bibinfo{journal}{Phys. Rev. Lett.} \textbf{\bibinfo{volume}{87}},
  \bibinfo{pages}{192302} (\bibinfo{year}{2001}).

\bibitem[{\citenamefont{Krasnitz
  et~al.}(2003{\natexlab{a}})\citenamefont{Krasnitz, Nara, and
  Venugopalan}}]{Krasnitz:2003jw}
\bibinfo{author}{\bibfnamefont{A.}~\bibnamefont{Krasnitz}},
  \bibinfo{author}{\bibfnamefont{Y.}~\bibnamefont{Nara}}, \bibnamefont{and}
  \bibinfo{author}{\bibfnamefont{R.}~\bibnamefont{Venugopalan}},
  \bibinfo{journal}{Nucl. Phys.} \textbf{\bibinfo{volume}{A727}},
  \bibinfo{pages}{427} (\bibinfo{year}{2003}{\natexlab{a}}).

\bibitem[{\citenamefont{Lappi}(2003)}]{Lappi:2003bi}
\bibinfo{author}{\bibfnamefont{T.}~\bibnamefont{Lappi}},
  \bibinfo{journal}{Phys. Rev.} \textbf{\bibinfo{volume}{C67}},
  \bibinfo{pages}{054903} (\bibinfo{year}{2003}).

\bibitem[{\citenamefont{Lappi}(2006)}]{Lappi:2006hq}
\bibinfo{author}{\bibfnamefont{T.}~\bibnamefont{Lappi}},
  \bibinfo{journal}{Phys. Lett.} \textbf{\bibinfo{volume}{B643}},
  \bibinfo{pages}{11} (\bibinfo{year}{2006}).

\bibitem[{\citenamefont{Krasnitz
  et~al.}(2003{\natexlab{b}})\citenamefont{Krasnitz, Nara, and
  Venugopalan}}]{Krasnitz:2002ng}
\bibinfo{author}{\bibfnamefont{A.}~\bibnamefont{Krasnitz}},
  \bibinfo{author}{\bibfnamefont{Y.}~\bibnamefont{Nara}}, \bibnamefont{and}
  \bibinfo{author}{\bibfnamefont{R.}~\bibnamefont{Venugopalan}},
  \bibinfo{journal}{Phys. Lett.} \textbf{\bibinfo{volume}{B554}},
  \bibinfo{pages}{21} (\bibinfo{year}{2003}{\natexlab{b}}).

\bibitem[{\citenamefont{Krasnitz
  et~al.}(2003{\natexlab{c}})\citenamefont{Krasnitz, Nara, and
  Venugopalan}}]{Krasnitz:2002mn}
\bibinfo{author}{\bibfnamefont{A.}~\bibnamefont{Krasnitz}},
  \bibinfo{author}{\bibfnamefont{Y.}~\bibnamefont{Nara}}, \bibnamefont{and}
  \bibinfo{author}{\bibfnamefont{R.}~\bibnamefont{Venugopalan}},
  \bibinfo{journal}{Nucl. Phys.} \textbf{\bibinfo{volume}{A717}},
  \bibinfo{pages}{268} (\bibinfo{year}{2003}{\natexlab{c}}).

\bibitem[{\citenamefont{Lappi and Venugopalan}(2006)}]{Lappi:2006xc}
\bibinfo{author}{\bibfnamefont{T.}~\bibnamefont{Lappi}} \bibnamefont{and}
  \bibinfo{author}{\bibfnamefont{R.}~\bibnamefont{Venugopalan}},
  \bibinfo{journal}{Phys. Rev.} \textbf{\bibinfo{volume}{C74}},
  \bibinfo{pages}{054905} (\bibinfo{year}{2006}).

\bibitem[{\citenamefont{Lappi and McLerran}(2006)}]{Lappi:2006fp}
\bibinfo{author}{\bibfnamefont{T.}~\bibnamefont{Lappi}} \bibnamefont{and}
  \bibinfo{author}{\bibfnamefont{L.}~\bibnamefont{McLerran}},
  \bibinfo{journal}{Nucl. Phys.} \textbf{\bibinfo{volume}{A772}},
  \bibinfo{pages}{200} (\bibinfo{year}{2006}), \eprint{hep-ph/0602189}.

\bibitem[{\citenamefont{Dumitru et~al.}(2013)\citenamefont{Dumitru, Nara, and
  Petreska}}]{Dumitru:2013koh}
\bibinfo{author}{\bibfnamefont{A.}~\bibnamefont{Dumitru}},
  \bibinfo{author}{\bibfnamefont{Y.}~\bibnamefont{Nara}}, \bibnamefont{and}
  \bibinfo{author}{\bibfnamefont{E.}~\bibnamefont{Petreska}},
  \bibinfo{journal}{Phys. Rev.} \textbf{\bibinfo{volume}{D88}},
  \bibinfo{pages}{054016} (\bibinfo{year}{2013}), \eprint{1302.2064}.

\bibitem[{\citenamefont{Dumitru et~al.}(2014)\citenamefont{Dumitru, Lappi, and
  Nara}}]{Dumitru:2014nka}
\bibinfo{author}{\bibfnamefont{A.}~\bibnamefont{Dumitru}},
  \bibinfo{author}{\bibfnamefont{T.}~\bibnamefont{Lappi}}, \bibnamefont{and}
  \bibinfo{author}{\bibfnamefont{Y.}~\bibnamefont{Nara}},
  \bibinfo{journal}{Phys. Lett.} \textbf{\bibinfo{volume}{B734}},
  \bibinfo{pages}{7} (\bibinfo{year}{2014}), \eprint{1401.4124}.

\bibitem[{\citenamefont{Bartels et~al.}(2002)\citenamefont{Bartels,
  Golec-Biernat, and Kowalski}}]{Bartels:2002cj}
\bibinfo{author}{\bibfnamefont{J.}~\bibnamefont{Bartels}},
  \bibinfo{author}{\bibfnamefont{K.~J.} \bibnamefont{Golec-Biernat}},
  \bibnamefont{and} \bibinfo{author}{\bibfnamefont{H.}~\bibnamefont{Kowalski}},
  \bibinfo{journal}{Phys. Rev.} \textbf{\bibinfo{volume}{D66}},
  \bibinfo{pages}{014001} (\bibinfo{year}{2002}).

\bibitem[{\citenamefont{Kowalski and Teaney}(2003)}]{Kowalski:2003hm}
\bibinfo{author}{\bibfnamefont{H.}~\bibnamefont{Kowalski}} \bibnamefont{and}
  \bibinfo{author}{\bibfnamefont{D.}~\bibnamefont{Teaney}},
  \bibinfo{journal}{Phys. Rev.} \textbf{\bibinfo{volume}{D68}},
  \bibinfo{pages}{114005} (\bibinfo{year}{2003}).

\bibitem[{\citenamefont{Kowalski et~al.}(2008)\citenamefont{Kowalski, Lappi,
  and Venugopalan}}]{Kowalski:2007rw}
\bibinfo{author}{\bibfnamefont{H.}~\bibnamefont{Kowalski}},
  \bibinfo{author}{\bibfnamefont{T.}~\bibnamefont{Lappi}}, \bibnamefont{and}
  \bibinfo{author}{\bibfnamefont{R.}~\bibnamefont{Venugopalan}},
  \bibinfo{journal}{Phys. Rev. Lett.} \textbf{\bibinfo{volume}{100}},
  \bibinfo{pages}{022303} (\bibinfo{year}{2008}).

\bibitem[{\citenamefont{Kowalski et~al.}(2006)\citenamefont{Kowalski, Motyka,
  and Watt}}]{Kowalski:2006hc}
\bibinfo{author}{\bibfnamefont{H.}~\bibnamefont{Kowalski}},
  \bibinfo{author}{\bibfnamefont{L.}~\bibnamefont{Motyka}}, \bibnamefont{and}
  \bibinfo{author}{\bibfnamefont{G.}~\bibnamefont{Watt}},
  \bibinfo{journal}{Phys. Rev.} \textbf{\bibinfo{volume}{D74}},
  \bibinfo{pages}{074016} (\bibinfo{year}{2006}).

\bibitem[{\citenamefont{Schenke
  et~al.}(2014{\natexlab{a}})\citenamefont{Schenke, Tribedy, and
  Venugopalan}}]{Schenke:2013dpa}
\bibinfo{author}{\bibfnamefont{B.}~\bibnamefont{Schenke}},
  \bibinfo{author}{\bibfnamefont{P.}~\bibnamefont{Tribedy}}, \bibnamefont{and}
  \bibinfo{author}{\bibfnamefont{R.}~\bibnamefont{Venugopalan}},
  \bibinfo{journal}{Phys. Rev.} \textbf{\bibinfo{volume}{C89}},
  \bibinfo{pages}{024901} (\bibinfo{year}{2014}{\natexlab{a}}),
  \eprint{1311.3636}.

\bibitem[{\citenamefont{Rezaeian et~al.}(2013)\citenamefont{Rezaeian, Siddikov,
  Van~de Klundert, and Venugopalan}}]{Rezaeian:2012ji}
\bibinfo{author}{\bibfnamefont{A.~H.} \bibnamefont{Rezaeian}},
  \bibinfo{author}{\bibfnamefont{M.}~\bibnamefont{Siddikov}},
  \bibinfo{author}{\bibfnamefont{M.}~\bibnamefont{Van~de Klundert}},
  \bibnamefont{and}
  \bibinfo{author}{\bibfnamefont{R.}~\bibnamefont{Venugopalan}},
  \bibinfo{journal}{Phys.Rev.} \textbf{\bibinfo{volume}{D87}},
  \bibinfo{pages}{034002} (\bibinfo{year}{2013}), \eprint{1212.2974}.

\bibitem[{\citenamefont{Lappi}(2008)}]{Lappi:2007ku}
\bibinfo{author}{\bibfnamefont{T.}~\bibnamefont{Lappi}}, \bibinfo{journal}{Eur.
  Phys. J.} \textbf{\bibinfo{volume}{C55}}, \bibinfo{pages}{285}
  (\bibinfo{year}{2008}).

\bibitem[{\citenamefont{Gelis et~al.}(2009{\natexlab{a}})\citenamefont{Gelis,
  Lappi, and McLerran}}]{Gelis:2009wh}
\bibinfo{author}{\bibfnamefont{F.}~\bibnamefont{Gelis}},
  \bibinfo{author}{\bibfnamefont{T.}~\bibnamefont{Lappi}}, \bibnamefont{and}
  \bibinfo{author}{\bibfnamefont{L.}~\bibnamefont{McLerran}},
  \bibinfo{journal}{Nucl. Phys.} \textbf{\bibinfo{volume}{A828}},
  \bibinfo{pages}{149} (\bibinfo{year}{2009}{\natexlab{a}}).

\bibitem[{\citenamefont{Abelev et~al.}(2009)}]{Abelev:2008ez}
\bibinfo{author}{\bibfnamefont{B.~I.} \bibnamefont{Abelev}}
  \bibnamefont{et~al.} (\bibinfo{collaboration}{STAR}), \bibinfo{journal}{Phys.
  Rev.} \textbf{\bibinfo{volume}{C79}}, \bibinfo{pages}{034909}
  (\bibinfo{year}{2009}), \eprint{0808.2041}.

\bibitem[{\citenamefont{Cooper and Frye}(1974)}]{Cooper:1974mv}
\bibinfo{author}{\bibfnamefont{F.}~\bibnamefont{Cooper}} \bibnamefont{and}
  \bibinfo{author}{\bibfnamefont{G.}~\bibnamefont{Frye}},
  \bibinfo{journal}{Phys. Rev.} \textbf{\bibinfo{volume}{D10}},
  \bibinfo{pages}{186} (\bibinfo{year}{1974}).

\bibitem[{\citenamefont{Qiu et~al.}(2012)\citenamefont{Qiu, Shen, and
  Heinz}}]{Qiu:2011hf}
\bibinfo{author}{\bibfnamefont{Z.}~\bibnamefont{Qiu}},
  \bibinfo{author}{\bibfnamefont{C.}~\bibnamefont{Shen}}, \bibnamefont{and}
  \bibinfo{author}{\bibfnamefont{U.~W.} \bibnamefont{Heinz}},
  \bibinfo{journal}{Phys. Lett.} \textbf{\bibinfo{volume}{B707}},
  \bibinfo{pages}{151} (\bibinfo{year}{2012}).

\bibitem[{\citenamefont{Aad et~al.}(2012)}]{ATLAS:2012at}
\bibinfo{author}{\bibfnamefont{G.}~\bibnamefont{Aad}} \bibnamefont{et~al.}
  (\bibinfo{collaboration}{ATLAS Collaboration}), \bibinfo{journal}{Phys.Rev.}
  \textbf{\bibinfo{volume}{C86}}, \bibinfo{pages}{014907}
  (\bibinfo{year}{2012}).

\bibitem[{\citenamefont{Aamodt et~al.}(2011)}]{ALICE:2011ab}
\bibinfo{author}{\bibfnamefont{K.}~\bibnamefont{Aamodt}} \bibnamefont{et~al.}
  (\bibinfo{collaboration}{ALICE Collaboration}),
  \bibinfo{journal}{Phys.Rev.Lett.} \textbf{\bibinfo{volume}{107}},
  \bibinfo{pages}{032301} (\bibinfo{year}{2011}).

\bibitem[{\citenamefont{Schenke and Venugopalan}(2014)}]{Schenke:2014zha}
\bibinfo{author}{\bibfnamefont{B.}~\bibnamefont{Schenke}} \bibnamefont{and}
  \bibinfo{author}{\bibfnamefont{R.}~\bibnamefont{Venugopalan}},
  \bibinfo{journal}{Phys.Rev.Lett.} \textbf{\bibinfo{volume}{113}},
  \bibinfo{pages}{102301} (\bibinfo{year}{2014}), \eprint{1405.3605}.

\bibitem[{\citenamefont{Adamczyk et~al.}(2015)}]{Adamczyk:2015obl}
\bibinfo{author}{\bibfnamefont{L.}~\bibnamefont{Adamczyk}} \bibnamefont{et~al.}
  (\bibinfo{collaboration}{STAR}), \bibinfo{journal}{Phys. Rev. Lett.}
  \textbf{\bibinfo{volume}{115}}, \bibinfo{pages}{222301}
  (\bibinfo{year}{2015}), \eprint{1505.07812}.

\bibitem[{\citenamefont{Schenke
  et~al.}(2014{\natexlab{b}})\citenamefont{Schenke, Tribedy, and
  Venugopalan}}]{Schenke:2014tga}
\bibinfo{author}{\bibfnamefont{B.}~\bibnamefont{Schenke}},
  \bibinfo{author}{\bibfnamefont{P.}~\bibnamefont{Tribedy}}, \bibnamefont{and}
  \bibinfo{author}{\bibfnamefont{R.}~\bibnamefont{Venugopalan}},
  \bibinfo{journal}{Phys. Rev.} \textbf{\bibinfo{volume}{C89}},
  \bibinfo{pages}{064908} (\bibinfo{year}{2014}{\natexlab{b}}),
  \eprint{1403.2232}.

\bibitem[{\citenamefont{Gelis and Venugopalan}(2006)}]{Gelis:2006yv}
\bibinfo{author}{\bibfnamefont{F.}~\bibnamefont{Gelis}} \bibnamefont{and}
  \bibinfo{author}{\bibfnamefont{R.}~\bibnamefont{Venugopalan}},
  \bibinfo{journal}{Nucl. Phys.} \textbf{\bibinfo{volume}{A776}},
  \bibinfo{pages}{135} (\bibinfo{year}{2006}), \eprint{hep-ph/0601209}.

\bibitem[{\citenamefont{Gelis et~al.}(2007)\citenamefont{Gelis, Lappi, and
  Venugopalan}}]{Gelis:2007kn}
\bibinfo{author}{\bibfnamefont{F.}~\bibnamefont{Gelis}},
  \bibinfo{author}{\bibfnamefont{T.}~\bibnamefont{Lappi}}, \bibnamefont{and}
  \bibinfo{author}{\bibfnamefont{R.}~\bibnamefont{Venugopalan}},
  \bibinfo{journal}{Int. J. Mod. Phys.} \textbf{\bibinfo{volume}{E16}},
  \bibinfo{pages}{2595} (\bibinfo{year}{2007}), \eprint{0708.0047}.

\bibitem[{\citenamefont{Epelbaum and Gelis}(2011)}]{Epelbaum:2011pc}
\bibinfo{author}{\bibfnamefont{T.}~\bibnamefont{Epelbaum}} \bibnamefont{and}
  \bibinfo{author}{\bibfnamefont{F.}~\bibnamefont{Gelis}},
  \bibinfo{journal}{Nucl. Phys.} \textbf{\bibinfo{volume}{A872}},
  \bibinfo{pages}{210} (\bibinfo{year}{2011}), \eprint{1107.0668}.

\bibitem[{\citenamefont{Gelis et~al.}(2008{\natexlab{a}})\citenamefont{Gelis,
  Lappi, and Venugopalan}}]{Gelis:2008rw}
\bibinfo{author}{\bibfnamefont{F.}~\bibnamefont{Gelis}},
  \bibinfo{author}{\bibfnamefont{T.}~\bibnamefont{Lappi}}, \bibnamefont{and}
  \bibinfo{author}{\bibfnamefont{R.}~\bibnamefont{Venugopalan}},
  \bibinfo{journal}{Phys. Rev.} \textbf{\bibinfo{volume}{D78}},
  \bibinfo{pages}{054019} (\bibinfo{year}{2008}{\natexlab{a}}),
  \eprint{0804.2630}.

\bibitem[{\citenamefont{Gelis and Laidet}(2013)}]{Gelis:2012ct}
\bibinfo{author}{\bibfnamefont{F.}~\bibnamefont{Gelis}} \bibnamefont{and}
  \bibinfo{author}{\bibfnamefont{J.~u.} \bibnamefont{Laidet}},
  \bibinfo{journal}{Phys. Rev.} \textbf{\bibinfo{volume}{D87}},
  \bibinfo{pages}{045019} (\bibinfo{year}{2013}), \eprint{1211.1191}.

\bibitem[{\citenamefont{Gelis et~al.}(2008{\natexlab{b}})\citenamefont{Gelis,
  Lappi, and Venugopalan}}]{Gelis:2008ad}
\bibinfo{author}{\bibfnamefont{F.}~\bibnamefont{Gelis}},
  \bibinfo{author}{\bibfnamefont{T.}~\bibnamefont{Lappi}}, \bibnamefont{and}
  \bibinfo{author}{\bibfnamefont{R.}~\bibnamefont{Venugopalan}},
  \bibinfo{journal}{Phys. Rev.} \textbf{\bibinfo{volume}{D78}},
  \bibinfo{pages}{054020} (\bibinfo{year}{2008}{\natexlab{b}}),
  \eprint{0807.1306}.

\bibitem[{\citenamefont{Gelis et~al.}(2009{\natexlab{b}})\citenamefont{Gelis,
  Lappi, and Venugopalan}}]{Gelis:2008sz}
\bibinfo{author}{\bibfnamefont{F.}~\bibnamefont{Gelis}},
  \bibinfo{author}{\bibfnamefont{T.}~\bibnamefont{Lappi}}, \bibnamefont{and}
  \bibinfo{author}{\bibfnamefont{R.}~\bibnamefont{Venugopalan}},
  \bibinfo{journal}{Phys. Rev.} \textbf{\bibinfo{volume}{D79}},
  \bibinfo{pages}{094017} (\bibinfo{year}{2009}{\natexlab{b}}),
  \eprint{0810.4829}.

\bibitem[{\citenamefont{Balitsky}(1996)}]{Balitsky:1995ub}
\bibinfo{author}{\bibfnamefont{I.}~\bibnamefont{Balitsky}},
  \bibinfo{journal}{Nucl. Phys.} \textbf{\bibinfo{volume}{B463}},
  \bibinfo{pages}{99} (\bibinfo{year}{1996}), \eprint{hep-ph/9509348}.

\bibitem[{\citenamefont{Jalilian-Marian
  et~al.}(1997{\natexlab{a}})\citenamefont{Jalilian-Marian, Kovner, McLerran,
  and Weigert}}]{JalilianMarian:1996xn}
\bibinfo{author}{\bibfnamefont{J.}~\bibnamefont{Jalilian-Marian}},
  \bibinfo{author}{\bibfnamefont{A.}~\bibnamefont{Kovner}},
  \bibinfo{author}{\bibfnamefont{L.~D.} \bibnamefont{McLerran}},
  \bibnamefont{and} \bibinfo{author}{\bibfnamefont{H.}~\bibnamefont{Weigert}},
  \bibinfo{journal}{Phys. Rev.} \textbf{\bibinfo{volume}{D55}},
  \bibinfo{pages}{5414} (\bibinfo{year}{1997}{\natexlab{a}}),
  \eprint{hep-ph/9606337}.

\bibitem[{\citenamefont{Jalilian-Marian
  et~al.}(1997{\natexlab{b}})\citenamefont{Jalilian-Marian, Kovner, Leonidov,
  and Weigert}}]{JalilianMarian:1997jx}
\bibinfo{author}{\bibfnamefont{J.}~\bibnamefont{Jalilian-Marian}},
  \bibinfo{author}{\bibfnamefont{A.}~\bibnamefont{Kovner}},
  \bibinfo{author}{\bibfnamefont{A.}~\bibnamefont{Leonidov}}, \bibnamefont{and}
  \bibinfo{author}{\bibfnamefont{H.}~\bibnamefont{Weigert}},
  \bibinfo{journal}{Nucl. Phys.} \textbf{\bibinfo{volume}{B504}},
  \bibinfo{pages}{415} (\bibinfo{year}{1997}{\natexlab{b}}),
  \eprint{hep-ph/9701284}.

\bibitem[{\citenamefont{Jalilian-Marian
  et~al.}(1998)\citenamefont{Jalilian-Marian, Kovner, Leonidov, and
  Weigert}}]{JalilianMarian:1997gr}
\bibinfo{author}{\bibfnamefont{J.}~\bibnamefont{Jalilian-Marian}},
  \bibinfo{author}{\bibfnamefont{A.}~\bibnamefont{Kovner}},
  \bibinfo{author}{\bibfnamefont{A.}~\bibnamefont{Leonidov}}, \bibnamefont{and}
  \bibinfo{author}{\bibfnamefont{H.}~\bibnamefont{Weigert}},
  \bibinfo{journal}{Phys. Rev.} \textbf{\bibinfo{volume}{D59}},
  \bibinfo{pages}{014014} (\bibinfo{year}{1998}), \eprint{hep-ph/9706377}.

\bibitem[{\citenamefont{Jalilian-Marian
  et~al.}(1999{\natexlab{a}})\citenamefont{Jalilian-Marian, Kovner, and
  Weigert}}]{JalilianMarian:1997dw}
\bibinfo{author}{\bibfnamefont{J.}~\bibnamefont{Jalilian-Marian}},
  \bibinfo{author}{\bibfnamefont{A.}~\bibnamefont{Kovner}}, \bibnamefont{and}
  \bibinfo{author}{\bibfnamefont{H.}~\bibnamefont{Weigert}},
  \bibinfo{journal}{Phys. Rev.} \textbf{\bibinfo{volume}{D59}},
  \bibinfo{pages}{014015} (\bibinfo{year}{1999}{\natexlab{a}}),
  \eprint{hep-ph/9709432}.

\bibitem[{\citenamefont{Jalilian-Marian
  et~al.}(1999{\natexlab{b}})\citenamefont{Jalilian-Marian, Kovner, Leonidov,
  and Weigert}}]{JalilianMarian:1998cb}
\bibinfo{author}{\bibfnamefont{J.}~\bibnamefont{Jalilian-Marian}},
  \bibinfo{author}{\bibfnamefont{A.}~\bibnamefont{Kovner}},
  \bibinfo{author}{\bibfnamefont{A.}~\bibnamefont{Leonidov}}, \bibnamefont{and}
  \bibinfo{author}{\bibfnamefont{H.}~\bibnamefont{Weigert}},
  \bibinfo{journal}{Phys. Rev.} \textbf{\bibinfo{volume}{D59}},
  \bibinfo{pages}{034007} (\bibinfo{year}{1999}{\natexlab{b}}),
  \bibinfo{note}{[Erratum: Phys. Rev.D59,099903(1999)]},
  \eprint{hep-ph/9807462}.

\bibitem[{\citenamefont{Iancu et~al.}(2001{\natexlab{a}})\citenamefont{Iancu,
  Leonidov, and McLerran}}]{Iancu:2000hn}
\bibinfo{author}{\bibfnamefont{E.}~\bibnamefont{Iancu}},
  \bibinfo{author}{\bibfnamefont{A.}~\bibnamefont{Leonidov}}, \bibnamefont{and}
  \bibinfo{author}{\bibfnamefont{L.~D.} \bibnamefont{McLerran}},
  \bibinfo{journal}{Nucl. Phys.} \textbf{\bibinfo{volume}{A692}},
  \bibinfo{pages}{583} (\bibinfo{year}{2001}{\natexlab{a}}),
  \eprint{hep-ph/0011241}.

\bibitem[{\citenamefont{Iancu et~al.}(2001{\natexlab{b}})\citenamefont{Iancu,
  Leonidov, and McLerran}}]{Iancu:2001ad}
\bibinfo{author}{\bibfnamefont{E.}~\bibnamefont{Iancu}},
  \bibinfo{author}{\bibfnamefont{A.}~\bibnamefont{Leonidov}}, \bibnamefont{and}
  \bibinfo{author}{\bibfnamefont{L.~D.} \bibnamefont{McLerran}},
  \bibinfo{journal}{Phys. Lett.} \textbf{\bibinfo{volume}{B510}},
  \bibinfo{pages}{133} (\bibinfo{year}{2001}{\natexlab{b}}),
  \eprint{hep-ph/0102009}.

\bibitem[{\citenamefont{Ferreiro et~al.}(2002)\citenamefont{Ferreiro, Iancu,
  Leonidov, and McLerran}}]{Ferreiro:2001qy}
\bibinfo{author}{\bibfnamefont{E.}~\bibnamefont{Ferreiro}},
  \bibinfo{author}{\bibfnamefont{E.}~\bibnamefont{Iancu}},
  \bibinfo{author}{\bibfnamefont{A.}~\bibnamefont{Leonidov}}, \bibnamefont{and}
  \bibinfo{author}{\bibfnamefont{L.}~\bibnamefont{McLerran}},
  \bibinfo{journal}{Nucl. Phys.} \textbf{\bibinfo{volume}{A703}},
  \bibinfo{pages}{489} (\bibinfo{year}{2002}), \eprint{hep-ph/0109115}.

\bibitem[{\citenamefont{Balitsky and Chirilli}(2008)}]{Balitsky:2008zza}
\bibinfo{author}{\bibfnamefont{I.}~\bibnamefont{Balitsky}} \bibnamefont{and}
  \bibinfo{author}{\bibfnamefont{G.~A.} \bibnamefont{Chirilli}},
  \bibinfo{journal}{Phys. Rev.} \textbf{\bibinfo{volume}{D77}},
  \bibinfo{pages}{014019} (\bibinfo{year}{2008}), \eprint{0710.4330}.

\bibitem[{\citenamefont{Balitsky and Chirilli}(2009)}]{Balitsky:2009xg}
\bibinfo{author}{\bibfnamefont{I.}~\bibnamefont{Balitsky}} \bibnamefont{and}
  \bibinfo{author}{\bibfnamefont{G.~A.} \bibnamefont{Chirilli}},
  \bibinfo{journal}{Nucl. Phys.} \textbf{\bibinfo{volume}{B822}},
  \bibinfo{pages}{45} (\bibinfo{year}{2009}), \eprint{0903.5326}.

\bibitem[{\citenamefont{Balitsky and Chirilli}(2013)}]{Balitsky:2013fea}
\bibinfo{author}{\bibfnamefont{I.}~\bibnamefont{Balitsky}} \bibnamefont{and}
  \bibinfo{author}{\bibfnamefont{G.~A.} \bibnamefont{Chirilli}},
  \bibinfo{journal}{Phys. Rev.} \textbf{\bibinfo{volume}{D88}},
  \bibinfo{pages}{111501} (\bibinfo{year}{2013}), \eprint{1309.7644}.

\bibitem[{\citenamefont{Grabovsky}(2013)}]{Grabovsky:2013mba}
\bibinfo{author}{\bibfnamefont{A.~V.} \bibnamefont{Grabovsky}},
  \bibinfo{journal}{JHEP} \textbf{\bibinfo{volume}{09}}, \bibinfo{pages}{141}
  (\bibinfo{year}{2013}), \eprint{1307.5414}.

\bibitem[{\citenamefont{Kovner et~al.}(2014{\natexlab{a}})\citenamefont{Kovner,
  Lublinsky, and Mulian}}]{Kovner:2013ona}
\bibinfo{author}{\bibfnamefont{A.}~\bibnamefont{Kovner}},
  \bibinfo{author}{\bibfnamefont{M.}~\bibnamefont{Lublinsky}},
  \bibnamefont{and} \bibinfo{author}{\bibfnamefont{Y.}~\bibnamefont{Mulian}},
  \bibinfo{journal}{Phys. Rev.} \textbf{\bibinfo{volume}{D89}},
  \bibinfo{pages}{061704} (\bibinfo{year}{2014}{\natexlab{a}}),
  \eprint{1310.0378}.

\bibitem[{\citenamefont{Kovner et~al.}(2014{\natexlab{b}})\citenamefont{Kovner,
  Lublinsky, and Mulian}}]{Kovner:2014xia}
\bibinfo{author}{\bibfnamefont{A.}~\bibnamefont{Kovner}},
  \bibinfo{author}{\bibfnamefont{M.}~\bibnamefont{Lublinsky}},
  \bibnamefont{and} \bibinfo{author}{\bibfnamefont{Y.}~\bibnamefont{Mulian}},
  \bibinfo{journal}{JHEP} \textbf{\bibinfo{volume}{04}}, \bibinfo{pages}{030}
  (\bibinfo{year}{2014}{\natexlab{b}}), \eprint{1401.0374}.

\bibitem[{\citenamefont{Kovner et~al.}(2014{\natexlab{c}})\citenamefont{Kovner,
  Lublinsky, and Mulian}}]{Kovner:2014lca}
\bibinfo{author}{\bibfnamefont{A.}~\bibnamefont{Kovner}},
  \bibinfo{author}{\bibfnamefont{M.}~\bibnamefont{Lublinsky}},
  \bibnamefont{and} \bibinfo{author}{\bibfnamefont{Y.}~\bibnamefont{Mulian}},
  \bibinfo{journal}{JHEP} \textbf{\bibinfo{volume}{08}}, \bibinfo{pages}{114}
  (\bibinfo{year}{2014}{\natexlab{c}}), \eprint{1405.0418}.

\bibitem[{\citenamefont{Blaizot et~al.}(2003)\citenamefont{Blaizot, Iancu, and
  Weigert}}]{Blaizot:2002np}
\bibinfo{author}{\bibfnamefont{J.-P.} \bibnamefont{Blaizot}},
  \bibinfo{author}{\bibfnamefont{E.}~\bibnamefont{Iancu}}, \bibnamefont{and}
  \bibinfo{author}{\bibfnamefont{H.}~\bibnamefont{Weigert}},
  \bibinfo{journal}{Nucl. Phys.} \textbf{\bibinfo{volume}{A713}},
  \bibinfo{pages}{441} (\bibinfo{year}{2003}), \eprint{hep-ph/0206279}.

\bibitem[{\citenamefont{Rummukainen and Weigert}(2004)}]{Rummukainen:2003ns}
\bibinfo{author}{\bibfnamefont{K.}~\bibnamefont{Rummukainen}} \bibnamefont{and}
  \bibinfo{author}{\bibfnamefont{H.}~\bibnamefont{Weigert}},
  \bibinfo{journal}{Nucl. Phys.} \textbf{\bibinfo{volume}{A739}},
  \bibinfo{pages}{183} (\bibinfo{year}{2004}), \eprint{hep-ph/0309306}.

\bibitem[{\citenamefont{Dumitru et~al.}(2011)\citenamefont{Dumitru,
  Jalilian-Marian, Lappi, Schenke, and Venugopalan}}]{Dumitru:2011vk}
\bibinfo{author}{\bibfnamefont{A.}~\bibnamefont{Dumitru}},
  \bibinfo{author}{\bibfnamefont{J.}~\bibnamefont{Jalilian-Marian}},
  \bibinfo{author}{\bibfnamefont{T.}~\bibnamefont{Lappi}},
  \bibinfo{author}{\bibfnamefont{B.}~\bibnamefont{Schenke}}, \bibnamefont{and}
  \bibinfo{author}{\bibfnamefont{R.}~\bibnamefont{Venugopalan}},
  \bibinfo{journal}{Phys. Lett.} \textbf{\bibinfo{volume}{B706}},
  \bibinfo{pages}{219} (\bibinfo{year}{2011}), \eprint{1108.4764}.

\bibitem[{\citenamefont{Kovchegov}(1999)}]{Kovchegov:1999yj}
\bibinfo{author}{\bibfnamefont{Y.~V.} \bibnamefont{Kovchegov}},
  \bibinfo{journal}{Phys. Rev.} \textbf{\bibinfo{volume}{D60}},
  \bibinfo{pages}{034008} (\bibinfo{year}{1999}), \eprint{hep-ph/9901281}.

\bibitem[{\citenamefont{Balitsky}(2007)}]{Balitsky:2006wa}
\bibinfo{author}{\bibfnamefont{I.}~\bibnamefont{Balitsky}},
  \bibinfo{journal}{Phys. Rev.} \textbf{\bibinfo{volume}{D75}},
  \bibinfo{pages}{014001} (\bibinfo{year}{2007}), \eprint{hep-ph/0609105}.

\bibitem[{\citenamefont{Kovchegov and Weigert}(2007)}]{Kovchegov:2006vj}
\bibinfo{author}{\bibfnamefont{Y.~V.} \bibnamefont{Kovchegov}}
  \bibnamefont{and} \bibinfo{author}{\bibfnamefont{H.}~\bibnamefont{Weigert}},
  \bibinfo{journal}{Nucl. Phys.} \textbf{\bibinfo{volume}{A784}},
  \bibinfo{pages}{188} (\bibinfo{year}{2007}), \eprint{hep-ph/0609090}.

\bibitem[{\citenamefont{Gardi et~al.}(2007)\citenamefont{Gardi, Kuokkanen,
  Rummukainen, and Weigert}}]{Gardi:2006rp}
\bibinfo{author}{\bibfnamefont{E.}~\bibnamefont{Gardi}},
  \bibinfo{author}{\bibfnamefont{J.}~\bibnamefont{Kuokkanen}},
  \bibinfo{author}{\bibfnamefont{K.}~\bibnamefont{Rummukainen}},
  \bibnamefont{and} \bibinfo{author}{\bibfnamefont{H.}~\bibnamefont{Weigert}},
  \bibinfo{journal}{Nucl. Phys.} \textbf{\bibinfo{volume}{A784}},
  \bibinfo{pages}{282} (\bibinfo{year}{2007}), \eprint{hep-ph/0609087}.

\bibitem[{\citenamefont{Epelbaum and
  Gelis}(2013{\natexlab{a}})}]{Epelbaum:2013waa}
\bibinfo{author}{\bibfnamefont{T.}~\bibnamefont{Epelbaum}} \bibnamefont{and}
  \bibinfo{author}{\bibfnamefont{F.}~\bibnamefont{Gelis}},
  \bibinfo{journal}{Phys. Rev.} \textbf{\bibinfo{volume}{D88}},
  \bibinfo{pages}{085015} (\bibinfo{year}{2013}{\natexlab{a}}),
  \eprint{1307.1765}.

\bibitem[{\citenamefont{Mrowczynski}(1993)}]{Mrowczynski:1993qm}
\bibinfo{author}{\bibfnamefont{S.}~\bibnamefont{Mrowczynski}},
  \bibinfo{journal}{Phys. Lett.} \textbf{\bibinfo{volume}{B314}},
  \bibinfo{pages}{118} (\bibinfo{year}{1993}).

\bibitem[{\citenamefont{Mrowczynski}(1997)}]{Mrowczynski:1996vh}
\bibinfo{author}{\bibfnamefont{S.}~\bibnamefont{Mrowczynski}},
  \bibinfo{journal}{Phys. Lett.} \textbf{\bibinfo{volume}{B393}},
  \bibinfo{pages}{26} (\bibinfo{year}{1997}), \eprint{hep-ph/9606442}.

\bibitem[{\citenamefont{Biro et~al.}(1994)\citenamefont{Biro, Gong, Muller, and
  Trayanov}}]{Biro:1993qc}
\bibinfo{author}{\bibfnamefont{T.~S.} \bibnamefont{Biro}},
  \bibinfo{author}{\bibfnamefont{C.}~\bibnamefont{Gong}},
  \bibinfo{author}{\bibfnamefont{B.}~\bibnamefont{Muller}}, \bibnamefont{and}
  \bibinfo{author}{\bibfnamefont{A.}~\bibnamefont{Trayanov}},
  \bibinfo{journal}{Int. J. Mod. Phys.} \textbf{\bibinfo{volume}{C5}},
  \bibinfo{pages}{113} (\bibinfo{year}{1994}), \eprint{nucl-th/9306002}.

\bibitem[{\citenamefont{Heinz et~al.}(1997)\citenamefont{Heinz, Hu, Leupold,
  Matinyan, and Muller}}]{Heinz:1996wx}
\bibinfo{author}{\bibfnamefont{U.~W.} \bibnamefont{Heinz}},
  \bibinfo{author}{\bibfnamefont{C.~R.} \bibnamefont{Hu}},
  \bibinfo{author}{\bibfnamefont{S.}~\bibnamefont{Leupold}},
  \bibinfo{author}{\bibfnamefont{S.~G.} \bibnamefont{Matinyan}},
  \bibnamefont{and} \bibinfo{author}{\bibfnamefont{B.}~\bibnamefont{Muller}},
  \bibinfo{journal}{Phys. Rev.} \textbf{\bibinfo{volume}{D55}},
  \bibinfo{pages}{2464} (\bibinfo{year}{1997}), \eprint{hep-th/9608181}.

\bibitem[{\citenamefont{Bolte et~al.}(2000)\citenamefont{Bolte, Muller, and
  Schafer}}]{Bolte:1999th}
\bibinfo{author}{\bibfnamefont{J.}~\bibnamefont{Bolte}},
  \bibinfo{author}{\bibfnamefont{B.}~\bibnamefont{Muller}}, \bibnamefont{and}
  \bibinfo{author}{\bibfnamefont{A.}~\bibnamefont{Schafer}},
  \bibinfo{journal}{Phys. Rev.} \textbf{\bibinfo{volume}{D61}},
  \bibinfo{pages}{054506} (\bibinfo{year}{2000}), \eprint{hep-lat/9906037}.

\bibitem[{\citenamefont{Romatschke and
  Venugopalan}(2006{\natexlab{a}})}]{Romatschke:2005pm}
\bibinfo{author}{\bibfnamefont{P.}~\bibnamefont{Romatschke}} \bibnamefont{and}
  \bibinfo{author}{\bibfnamefont{R.}~\bibnamefont{Venugopalan}},
  \bibinfo{journal}{Phys. Rev. Lett.} \textbf{\bibinfo{volume}{96}},
  \bibinfo{pages}{062302} (\bibinfo{year}{2006}{\natexlab{a}}),
  \eprint{hep-ph/0510121}.

\bibitem[{\citenamefont{Romatschke and
  Venugopalan}(2006{\natexlab{b}})}]{Romatschke:2005ag}
\bibinfo{author}{\bibfnamefont{P.}~\bibnamefont{Romatschke}} \bibnamefont{and}
  \bibinfo{author}{\bibfnamefont{R.}~\bibnamefont{Venugopalan}},
  \bibinfo{journal}{Eur. Phys. J.} \textbf{\bibinfo{volume}{A29}},
  \bibinfo{pages}{71} (\bibinfo{year}{2006}{\natexlab{b}}),
  \eprint{hep-ph/0510292}.

\bibitem[{\citenamefont{Romatschke and
  Venugopalan}(2006{\natexlab{c}})}]{Romatschke:2006nk}
\bibinfo{author}{\bibfnamefont{P.}~\bibnamefont{Romatschke}} \bibnamefont{and}
  \bibinfo{author}{\bibfnamefont{R.}~\bibnamefont{Venugopalan}},
  \bibinfo{journal}{Phys. Rev.} \textbf{\bibinfo{volume}{D74}},
  \bibinfo{pages}{045011} (\bibinfo{year}{2006}{\natexlab{c}}),
  \eprint{hep-ph/0605045}.

\bibitem[{\citenamefont{Fukushima}(2007)}]{Fukushima:2007ja}
\bibinfo{author}{\bibfnamefont{K.}~\bibnamefont{Fukushima}},
  \bibinfo{journal}{Phys. Rev.} \textbf{\bibinfo{volume}{C76}},
  \bibinfo{pages}{021902} (\bibinfo{year}{2007}), \bibinfo{note}{[Erratum:
  Phys. Rev.C77,029901(2007)]}, \eprint{0711.2634}.

\bibitem[{\citenamefont{Fujii and Itakura}(2008)}]{Fujii:2008dd}
\bibinfo{author}{\bibfnamefont{H.}~\bibnamefont{Fujii}} \bibnamefont{and}
  \bibinfo{author}{\bibfnamefont{K.}~\bibnamefont{Itakura}},
  \bibinfo{journal}{Nucl. Phys.} \textbf{\bibinfo{volume}{A809}},
  \bibinfo{pages}{88} (\bibinfo{year}{2008}), \eprint{0803.0410}.

\bibitem[{\citenamefont{Fujii et~al.}(2009)\citenamefont{Fujii, Itakura, and
  Iwazaki}}]{Fujii:2009kb}
\bibinfo{author}{\bibfnamefont{H.}~\bibnamefont{Fujii}},
  \bibinfo{author}{\bibfnamefont{K.}~\bibnamefont{Itakura}}, \bibnamefont{and}
  \bibinfo{author}{\bibfnamefont{A.}~\bibnamefont{Iwazaki}},
  \bibinfo{journal}{Nucl. Phys.} \textbf{\bibinfo{volume}{A828}},
  \bibinfo{pages}{178} (\bibinfo{year}{2009}), \eprint{0903.2930}.

\bibitem[{\citenamefont{Kunihiro et~al.}(2010)\citenamefont{Kunihiro, Muller,
  Ohnishi, Schafer, Takahashi, and Yamamoto}}]{Kunihiro:2010tg}
\bibinfo{author}{\bibfnamefont{T.}~\bibnamefont{Kunihiro}},
  \bibinfo{author}{\bibfnamefont{B.}~\bibnamefont{Muller}},
  \bibinfo{author}{\bibfnamefont{A.}~\bibnamefont{Ohnishi}},
  \bibinfo{author}{\bibfnamefont{A.}~\bibnamefont{Schafer}},
  \bibinfo{author}{\bibfnamefont{T.~T.} \bibnamefont{Takahashi}},
  \bibnamefont{and} \bibinfo{author}{\bibfnamefont{A.}~\bibnamefont{Yamamoto}},
  \bibinfo{journal}{Phys. Rev.} \textbf{\bibinfo{volume}{D82}},
  \bibinfo{pages}{114015} (\bibinfo{year}{2010}), \eprint{1008.1156}.

\bibitem[{\citenamefont{Fukushima and Gelis}(2012)}]{Fukushima:2011nq}
\bibinfo{author}{\bibfnamefont{K.}~\bibnamefont{Fukushima}} \bibnamefont{and}
  \bibinfo{author}{\bibfnamefont{F.}~\bibnamefont{Gelis}},
  \bibinfo{journal}{Nucl. Phys.} \textbf{\bibinfo{volume}{A874}},
  \bibinfo{pages}{108} (\bibinfo{year}{2012}), \eprint{1106.1396}.

\bibitem[{\citenamefont{Epelbaum and
  Gelis}(2013{\natexlab{b}})}]{Gelis:2013rba}
\bibinfo{author}{\bibfnamefont{T.}~\bibnamefont{Epelbaum}} \bibnamefont{and}
  \bibinfo{author}{\bibfnamefont{F.}~\bibnamefont{Gelis}},
  \bibinfo{journal}{Phys. Rev. Lett.} \textbf{\bibinfo{volume}{111}},
  \bibinfo{pages}{232301} (\bibinfo{year}{2013}{\natexlab{b}}),
  \eprint{1307.2214}.

\bibitem[{\citenamefont{Berges et~al.}(2014{\natexlab{a}})\citenamefont{Berges,
  Boguslavski, Schlichting, and Venugopalan}}]{Berges:2013lsa}
\bibinfo{author}{\bibfnamefont{J.}~\bibnamefont{Berges}},
  \bibinfo{author}{\bibfnamefont{K.}~\bibnamefont{Boguslavski}},
  \bibinfo{author}{\bibfnamefont{S.}~\bibnamefont{Schlichting}},
  \bibnamefont{and}
  \bibinfo{author}{\bibfnamefont{R.}~\bibnamefont{Venugopalan}},
  \bibinfo{journal}{JHEP} \textbf{\bibinfo{volume}{05}}, \bibinfo{pages}{054}
  (\bibinfo{year}{2014}{\natexlab{a}}), \eprint{1312.5216}.

\bibitem[{\citenamefont{Epelbaum
  et~al.}(2014{\natexlab{a}})\citenamefont{Epelbaum, Gelis, and
  Wu}}]{Epelbaum:2014yja}
\bibinfo{author}{\bibfnamefont{T.}~\bibnamefont{Epelbaum}},
  \bibinfo{author}{\bibfnamefont{F.}~\bibnamefont{Gelis}}, \bibnamefont{and}
  \bibinfo{author}{\bibfnamefont{B.}~\bibnamefont{Wu}}, \bibinfo{journal}{Phys.
  Rev.} \textbf{\bibinfo{volume}{D90}}, \bibinfo{pages}{065029}
  (\bibinfo{year}{2014}{\natexlab{a}}), \eprint{1402.0115}.

\bibitem[{\citenamefont{Epelbaum
  et~al.}(2014{\natexlab{b}})\citenamefont{Epelbaum, Gelis, Tanji, and
  Wu}}]{Epelbaum:2014mfa}
\bibinfo{author}{\bibfnamefont{T.}~\bibnamefont{Epelbaum}},
  \bibinfo{author}{\bibfnamefont{F.}~\bibnamefont{Gelis}},
  \bibinfo{author}{\bibfnamefont{N.}~\bibnamefont{Tanji}}, \bibnamefont{and}
  \bibinfo{author}{\bibfnamefont{B.}~\bibnamefont{Wu}}, \bibinfo{journal}{Phys.
  Rev.} \textbf{\bibinfo{volume}{D90}}, \bibinfo{pages}{125032}
  (\bibinfo{year}{2014}{\natexlab{b}}), \eprint{1409.0701}.

\bibitem[{\citenamefont{Berges et~al.}(2014{\natexlab{b}})\citenamefont{Berges,
  Boguslavski, Schlichting, and Venugopalan}}]{Berges:2013eia}
\bibinfo{author}{\bibfnamefont{J.}~\bibnamefont{Berges}},
  \bibinfo{author}{\bibfnamefont{K.}~\bibnamefont{Boguslavski}},
  \bibinfo{author}{\bibfnamefont{S.}~\bibnamefont{Schlichting}},
  \bibnamefont{and}
  \bibinfo{author}{\bibfnamefont{R.}~\bibnamefont{Venugopalan}},
  \bibinfo{journal}{Phys. Rev.} \textbf{\bibinfo{volume}{D89}},
  \bibinfo{pages}{074011} (\bibinfo{year}{2014}{\natexlab{b}}),
  \eprint{1303.5650}.

\bibitem[{\citenamefont{Berges et~al.}(2014{\natexlab{c}})\citenamefont{Berges,
  Boguslavski, Schlichting, and Venugopalan}}]{Berges:2013fga}
\bibinfo{author}{\bibfnamefont{J.}~\bibnamefont{Berges}},
  \bibinfo{author}{\bibfnamefont{K.}~\bibnamefont{Boguslavski}},
  \bibinfo{author}{\bibfnamefont{S.}~\bibnamefont{Schlichting}},
  \bibnamefont{and}
  \bibinfo{author}{\bibfnamefont{R.}~\bibnamefont{Venugopalan}},
  \bibinfo{journal}{Phys. Rev.} \textbf{\bibinfo{volume}{D89}},
  \bibinfo{pages}{114007} (\bibinfo{year}{2014}{\natexlab{c}}),
  \eprint{1311.3005}.

\bibitem[{\citenamefont{Berges et~al.}(2015)\citenamefont{Berges, Boguslavski,
  Schlichting, and Venugopalan}}]{Berges:2014bba}
\bibinfo{author}{\bibfnamefont{J.}~\bibnamefont{Berges}},
  \bibinfo{author}{\bibfnamefont{K.}~\bibnamefont{Boguslavski}},
  \bibinfo{author}{\bibfnamefont{S.}~\bibnamefont{Schlichting}},
  \bibnamefont{and}
  \bibinfo{author}{\bibfnamefont{R.}~\bibnamefont{Venugopalan}},
  \bibinfo{journal}{Phys. Rev. Lett.} \textbf{\bibinfo{volume}{114}},
  \bibinfo{pages}{061601} (\bibinfo{year}{2015}), \eprint{1408.1670}.

\bibitem[{\citenamefont{Luttinger and Ward}(1960)}]{Luttinger:1960ua}
\bibinfo{author}{\bibfnamefont{J.~M.} \bibnamefont{Luttinger}}
  \bibnamefont{and} \bibinfo{author}{\bibfnamefont{J.~C.} \bibnamefont{Ward}},
  \bibinfo{journal}{Phys. Rev.} \textbf{\bibinfo{volume}{118}},
  \bibinfo{pages}{1417} (\bibinfo{year}{1960}).

\bibitem[{\citenamefont{Aarts et~al.}(2002)\citenamefont{Aarts, Ahrensmeier,
  Baier, Berges, and Serreau}}]{Aarts:2002dj}
\bibinfo{author}{\bibfnamefont{G.}~\bibnamefont{Aarts}},
  \bibinfo{author}{\bibfnamefont{D.}~\bibnamefont{Ahrensmeier}},
  \bibinfo{author}{\bibfnamefont{R.}~\bibnamefont{Baier}},
  \bibinfo{author}{\bibfnamefont{J.}~\bibnamefont{Berges}}, \bibnamefont{and}
  \bibinfo{author}{\bibfnamefont{J.}~\bibnamefont{Serreau}},
  \bibinfo{journal}{Phys. Rev.} \textbf{\bibinfo{volume}{D66}},
  \bibinfo{pages}{045008} (\bibinfo{year}{2002}), \eprint{hep-ph/0201308}.

\bibitem[{\citenamefont{Calzetta and Hu}(2002)}]{Calzetta:2002ub}
\bibinfo{author}{\bibfnamefont{E.~A.} \bibnamefont{Calzetta}} \bibnamefont{and}
  \bibinfo{author}{\bibfnamefont{B.~L.} \bibnamefont{Hu}}
  (\bibinfo{year}{2002}), \eprint{hep-ph/0205271}.

\bibitem[{\citenamefont{Berges}(2004)}]{Berges:2004pu}
\bibinfo{author}{\bibfnamefont{J.}~\bibnamefont{Berges}},
  \bibinfo{journal}{Phys. Rev.} \textbf{\bibinfo{volume}{D70}},
  \bibinfo{pages}{105010} (\bibinfo{year}{2004}), \eprint{hep-ph/0401172}.

\bibitem[{\citenamefont{Berges}(2005)}]{Berges:2004yj}
\bibinfo{author}{\bibfnamefont{J.}~\bibnamefont{Berges}}, \bibinfo{journal}{AIP
  Conf. Proc.} \textbf{\bibinfo{volume}{739}}, \bibinfo{pages}{3}
  (\bibinfo{year}{2005}), \bibinfo{note}{[,3(2004)]}, \eprint{hep-ph/0409233}.

\bibitem[{\citenamefont{Baym and Kadanoff}(1961)}]{Baym:1961zz}
\bibinfo{author}{\bibfnamefont{G.}~\bibnamefont{Baym}} \bibnamefont{and}
  \bibinfo{author}{\bibfnamefont{L.~P.} \bibnamefont{Kadanoff}},
  \bibinfo{journal}{Phys. Rev.} \textbf{\bibinfo{volume}{124}},
  \bibinfo{pages}{287} (\bibinfo{year}{1961}).

\bibitem[{\citenamefont{Hatta and
  Nishiyama}(2012{\natexlab{a}})}]{Hatta:2011ky}
\bibinfo{author}{\bibfnamefont{Y.}~\bibnamefont{Hatta}} \bibnamefont{and}
  \bibinfo{author}{\bibfnamefont{A.}~\bibnamefont{Nishiyama}},
  \bibinfo{journal}{Nucl. Phys.} \textbf{\bibinfo{volume}{A873}},
  \bibinfo{pages}{47} (\bibinfo{year}{2012}{\natexlab{a}}), \eprint{1108.0818}.

\bibitem[{\citenamefont{Hatta and
  Nishiyama}(2012{\natexlab{b}})}]{Hatta:2012gq}
\bibinfo{author}{\bibfnamefont{Y.}~\bibnamefont{Hatta}} \bibnamefont{and}
  \bibinfo{author}{\bibfnamefont{A.}~\bibnamefont{Nishiyama}},
  \bibinfo{journal}{Phys. Rev.} \textbf{\bibinfo{volume}{D86}},
  \bibinfo{pages}{076002} (\bibinfo{year}{2012}{\natexlab{b}}),
  \eprint{1206.4743}.

\bibitem[{\citenamefont{Epelbaum et~al.}(2015)\citenamefont{Epelbaum, Gelis,
  Jeon, Moore, and Wu}}]{Epelbaum:2015vxa}
\bibinfo{author}{\bibfnamefont{T.}~\bibnamefont{Epelbaum}},
  \bibinfo{author}{\bibfnamefont{F.}~\bibnamefont{Gelis}},
  \bibinfo{author}{\bibfnamefont{S.}~\bibnamefont{Jeon}},
  \bibinfo{author}{\bibfnamefont{G.}~\bibnamefont{Moore}}, \bibnamefont{and}
  \bibinfo{author}{\bibfnamefont{B.}~\bibnamefont{Wu}}, \bibinfo{journal}{JHEP}
  \textbf{\bibinfo{volume}{09}}, \bibinfo{pages}{117} (\bibinfo{year}{2015}),
  \eprint{1506.05580}.

\bibitem[{\citenamefont{Bozek}(2012)}]{Bozek:2011if}
\bibinfo{author}{\bibfnamefont{P.}~\bibnamefont{Bozek}},
  \bibinfo{journal}{Phys.Rev.} \textbf{\bibinfo{volume}{C85}},
  \bibinfo{pages}{014911} (\bibinfo{year}{2012}), \eprint{1112.0915}.

\bibitem[{\citenamefont{Bozek and
  Broniowski}(2013{\natexlab{a}})}]{Bozek:2012gr}
\bibinfo{author}{\bibfnamefont{P.}~\bibnamefont{Bozek}} \bibnamefont{and}
  \bibinfo{author}{\bibfnamefont{W.}~\bibnamefont{Broniowski}},
  \bibinfo{journal}{Phys.Lett.} \textbf{\bibinfo{volume}{B718}},
  \bibinfo{pages}{1557} (\bibinfo{year}{2013}{\natexlab{a}}),
  \eprint{1211.0845}.

\bibitem[{\citenamefont{Bozek et~al.}(2013)\citenamefont{Bozek, Broniowski, and
  Torrieri}}]{Bozek:2013ska}
\bibinfo{author}{\bibfnamefont{P.}~\bibnamefont{Bozek}},
  \bibinfo{author}{\bibfnamefont{W.}~\bibnamefont{Broniowski}},
  \bibnamefont{and} \bibinfo{author}{\bibfnamefont{G.}~\bibnamefont{Torrieri}},
  \bibinfo{journal}{Phys.Rev.Lett.} \textbf{\bibinfo{volume}{111}},
  \bibinfo{pages}{172303} (\bibinfo{year}{2013}).

\bibitem[{\citenamefont{Werner et~al.}(2014{\natexlab{a}})\citenamefont{Werner,
  Guiot, Karpenko, and Pierog}}]{Werner:2013tya}
\bibinfo{author}{\bibfnamefont{K.}~\bibnamefont{Werner}},
  \bibinfo{author}{\bibfnamefont{B.}~\bibnamefont{Guiot}},
  \bibinfo{author}{\bibfnamefont{I.}~\bibnamefont{Karpenko}}, \bibnamefont{and}
  \bibinfo{author}{\bibfnamefont{T.}~\bibnamefont{Pierog}},
  \bibinfo{journal}{Phys. Rev.} \textbf{\bibinfo{volume}{C89}},
  \bibinfo{pages}{064903} (\bibinfo{year}{2014}{\natexlab{a}}),
  \eprint{1312.1233}.

\bibitem[{\citenamefont{Werner et~al.}(2014{\natexlab{b}})\citenamefont{Werner,
  Bleicher, Guiot, Karpenko, and Pierog}}]{Werner:2013ipa}
\bibinfo{author}{\bibfnamefont{K.}~\bibnamefont{Werner}},
  \bibinfo{author}{\bibfnamefont{M.}~\bibnamefont{Bleicher}},
  \bibinfo{author}{\bibfnamefont{B.}~\bibnamefont{Guiot}},
  \bibinfo{author}{\bibfnamefont{I.}~\bibnamefont{Karpenko}}, \bibnamefont{and}
  \bibinfo{author}{\bibfnamefont{T.}~\bibnamefont{Pierog}},
  \bibinfo{journal}{Phys. Rev. Lett.} \textbf{\bibinfo{volume}{112}},
  \bibinfo{pages}{232301} (\bibinfo{year}{2014}{\natexlab{b}}),
  \eprint{1307.4379}.

\bibitem[{\citenamefont{Bozek and
  Broniowski}(2013{\natexlab{b}})}]{Bozek:2013uha}
\bibinfo{author}{\bibfnamefont{P.}~\bibnamefont{Bozek}} \bibnamefont{and}
  \bibinfo{author}{\bibfnamefont{W.}~\bibnamefont{Broniowski}},
  \bibinfo{journal}{Phys.Rev.} \textbf{\bibinfo{volume}{C88}},
  \bibinfo{pages}{014903} (\bibinfo{year}{2013}{\natexlab{b}}),
  \eprint{1304.3044}.

\bibitem[{\citenamefont{Kozlov et~al.}(2014)\citenamefont{Kozlov, Luzum,
  Denicol, Jeon, and Gale}}]{Kozlov:2014fqa}
\bibinfo{author}{\bibfnamefont{I.}~\bibnamefont{Kozlov}},
  \bibinfo{author}{\bibfnamefont{M.}~\bibnamefont{Luzum}},
  \bibinfo{author}{\bibfnamefont{G.}~\bibnamefont{Denicol}},
  \bibinfo{author}{\bibfnamefont{S.}~\bibnamefont{Jeon}}, \bibnamefont{and}
  \bibinfo{author}{\bibfnamefont{C.}~\bibnamefont{Gale}}
  (\bibinfo{year}{2014}), \eprint{1405.3976}.

\bibitem[{\citenamefont{Schlichting and Schenke}(2014)}]{Schlichting:2014ipa}
\bibinfo{author}{\bibfnamefont{S.}~\bibnamefont{Schlichting}} \bibnamefont{and}
  \bibinfo{author}{\bibfnamefont{B.}~\bibnamefont{Schenke}},
  \bibinfo{journal}{Phys.Lett.} \textbf{\bibinfo{volume}{B739}},
  \bibinfo{pages}{313} (\bibinfo{year}{2014}), \eprint{1407.8458}.

\bibitem[{\citenamefont{Niemi and Denicol}(2014)}]{Niemi:2014wta}
\bibinfo{author}{\bibfnamefont{H.}~\bibnamefont{Niemi}} \bibnamefont{and}
  \bibinfo{author}{\bibfnamefont{G.~S.} \bibnamefont{Denicol}}
  (\bibinfo{year}{2014}), \eprint{1404.7327}.

\bibitem[{\citenamefont{Kovner and Lublinsky}(2011)}]{Kovner:2010xk}
\bibinfo{author}{\bibfnamefont{A.}~\bibnamefont{Kovner}} \bibnamefont{and}
  \bibinfo{author}{\bibfnamefont{M.}~\bibnamefont{Lublinsky}},
  \bibinfo{journal}{Phys. Rev.} \textbf{\bibinfo{volume}{D83}},
  \bibinfo{pages}{034017} (\bibinfo{year}{2011}), \eprint{1012.3398}.

\bibitem[{\citenamefont{Dusling and Venugopalan}(2013)}]{Dusling:2013qoz}
\bibinfo{author}{\bibfnamefont{K.}~\bibnamefont{Dusling}} \bibnamefont{and}
  \bibinfo{author}{\bibfnamefont{R.}~\bibnamefont{Venugopalan}},
  \bibinfo{journal}{Phys. Rev.} \textbf{\bibinfo{volume}{D87}},
  \bibinfo{pages}{094034} (\bibinfo{year}{2013}), \eprint{1302.7018}.

\bibitem[{\citenamefont{Dumitru et~al.}(2015)\citenamefont{Dumitru, McLerran,
  and Skokov}}]{Dumitru:2014yza}
\bibinfo{author}{\bibfnamefont{A.}~\bibnamefont{Dumitru}},
  \bibinfo{author}{\bibfnamefont{L.}~\bibnamefont{McLerran}}, \bibnamefont{and}
  \bibinfo{author}{\bibfnamefont{V.}~\bibnamefont{Skokov}},
  \bibinfo{journal}{Phys. Lett.} \textbf{\bibinfo{volume}{B743}},
  \bibinfo{pages}{134} (\bibinfo{year}{2015}), \eprint{1410.4844}.

\bibitem[{\citenamefont{Lappi}(2015)}]{Lappi:2015vha}
\bibinfo{author}{\bibfnamefont{T.}~\bibnamefont{Lappi}},
  \bibinfo{journal}{Phys. Lett.} \textbf{\bibinfo{volume}{B744}},
  \bibinfo{pages}{315} (\bibinfo{year}{2015}), \eprint{1501.05505}.

\bibitem[{\citenamefont{Lappi et~al.}(2015)\citenamefont{Lappi, Schenke,
  Schlichting, and Venugopalan}}]{Lappi:2015vta}
\bibinfo{author}{\bibfnamefont{T.}~\bibnamefont{Lappi}},
  \bibinfo{author}{\bibfnamefont{B.}~\bibnamefont{Schenke}},
  \bibinfo{author}{\bibfnamefont{S.}~\bibnamefont{Schlichting}},
  \bibnamefont{and}
  \bibinfo{author}{\bibfnamefont{R.}~\bibnamefont{Venugopalan}}
  (\bibinfo{year}{2015}), \eprint{1509.03499}.

\bibitem[{\citenamefont{Schenke et~al.}(2015)\citenamefont{Schenke,
  Schlichting, and Venugopalan}}]{Schenke:2015aqa}
\bibinfo{author}{\bibfnamefont{B.}~\bibnamefont{Schenke}},
  \bibinfo{author}{\bibfnamefont{S.}~\bibnamefont{Schlichting}},
  \bibnamefont{and}
  \bibinfo{author}{\bibfnamefont{R.}~\bibnamefont{Venugopalan}},
  \bibinfo{journal}{Phys. Lett.} \textbf{\bibinfo{volume}{B747}},
  \bibinfo{pages}{76} (\bibinfo{year}{2015}), \eprint{1502.01331}.

\bibitem[{\citenamefont{Dusling et~al.}(2015)\citenamefont{Dusling, Li, and
  Schenke}}]{Dusling:2015gta}
\bibinfo{author}{\bibfnamefont{K.}~\bibnamefont{Dusling}},
  \bibinfo{author}{\bibfnamefont{W.}~\bibnamefont{Li}}, \bibnamefont{and}
  \bibinfo{author}{\bibfnamefont{B.}~\bibnamefont{Schenke}}
  (\bibinfo{year}{2015}), \eprint{1509.07939}.

\end{thebibliography}

\end{document}